\documentclass[preprint,showpacs,aps,draft, floatfix]{revtex4}

\usepackage{graphicx}
\usepackage{dcolumn}
\usepackage{bm}

\input epsf 
\newcounter{subequation}[equation]
\makeatletter
\fussy
\def \with respect to { with respect to }

\def \susy { supersymmetry }
\def \susyq { supersymmetric }

\def\l {\lambda } 
 
\def \t {\theta }

\def\a {\alpha }
\def\dh {\partial }
\def \d {\delta }

\def \g {\gamma }

\def \b {\beta }

\def \s {\sigma }
\def \e {\epsilon }

\def \ud { {1 \over 2} }

\def \calr { {\cal R } }

\def \tchi  {{\tilde \chi }}
\def \Eslash {E \kern-.5em\slash}
\def \pslash {p \kern-.5em\slash}
\def \kslash {k \kern-.5em\slash}
\def \Dslash {D \kern-.5em\slash}
\def \hslash {h \kern-.5em\slash}
\def \dslash {\partial \kern-.5em\slash}
\def \vslash {v \kern-.5em\slash}

\makeatletter

	\def\thesubequation{\theequation\@alph\c@subequation}
	       \def\@subeqnnum{{\rm (\thesubequation)}}
	\def\slabel#1{\@bsphack\if@filesw {\let\thepage\relax
		 \xdef\@gtempa{\write\@auxout{\string
	\newlabel{#1}{{\thesubequation}{\thepage}}}}}\@gtempa
	   \if@nobreak \ifvmode\nobreak\fi\fi\fi\@esphack}
	       \def\subeqnarray{\stepcounter{equation}
       \let\@currentlabel=\theequation\global\c@subequation\@ne
			  \global\@eqnswtrue
	\global\@eqcnt\z@\tabskip\@centering\let\\=\@subeqncr
$$\halign to \displaywidth\bgroup\@eqnsel\hskip\@centering
  $\displaystyle\tabskip\z@{##}$&\global\@eqcnt\@ne
  \hskip 2\arraycolsep \hfil${##}$\hfil
  &\global\@eqcnt\tw@ \hskip 2\arraycolsep
  $\displaystyle\tabskip\z@{##}$\hfil
   \tabskip\@centering&\llap{##}\tabskip\z@\cr}
\def\endsubeqnarray{\@@subeqncr\egroup
                     $$\global\@ignoretrue}
\def\@subeqncr{{\ifnum0=`}\fi\@ifstar{\global\@eqpen\@M
    \@ysubeqncr}{\global\@eqpen\interdisplaylinepenalty \@ysubeqncr}}
\def\@ysubeqncr{\@ifnextchar [{\@xsubeqncr}{\@xsubeqncr[\z@]}}
\def\@xsubeqncr[#1]{\ifnum0=`{\fi}\@@subeqncr
   \noalign{\penalty\@eqpen\vskip\jot\vskip #1\relax}}
\def\@@subeqncr{\let\@tempa\relax
    \ifcase\@eqcnt \def\@tempa{& & &}\or \def\@tempa{& &}
      \else \def\@tempa{&}\fi
     \@tempa \if@eqnsw\@subeqnnum\refstepcounter{subequation}\fi
     \global\@eqnswtrue\global\@eqcnt\z@\cr}
\let\@ssubeqncr=\@subeqncr
\@namedef{subeqnarray*}{\def\@subeqncr{\nonumber\@ssubeqncr}\subeqnarray}
\@namedef{endsubeqnarray*}{\global\advance\c@equation\m@ne%
                           \nonumber\endsubeqnarray}

\@addtoreset{equation}{section}
\renewcommand{\theequation}{\thesection.\arabic{equation}}
\newcommand{\be}{\begin{equation}} 
\newcommand{\ee}{\end{equation}} 
\newcommand{\ba}{\begin{array}}
\newcommand{\ea}{\end{array}}
\newcommand{\bea}{\begin{eqnarray}} 
\newcommand{\eea}{\end{eqnarray}} 
\newcommand{\bsea}{\begin{subeqnarray}} 
\newcommand{\esea}{\end{subeqnarray}}

\begin{document}
\title{Implications of vacuum stability constraints on the nonminimal
supersymmetric standard model with lepton number violation} \author{
M. Chemtob} \email{marc.chemtob@cea.fr} \affiliation{ Service de
Physique Th\'eorique, CEA-Saclay F-91191 Gif-sur-Yvette Cedex France}
\author{ P.N. Pandita} \email{ ppandita@nehu.ac.in } \affiliation{
Department of Physics, North Eastern Hill University, Shillong 793
022, India}




\begin{abstract} \vspace*{10pt}
We carry out a detailed analysis of the scalar sector of the
nonminimal supersymmetric standard model with lepton number violation,
and study the constraints imposed on it by the stability of the
electroweak symmetry breaking vacuum.  The model contains a trilinear
lepton mumber violating term in the superpotential together with the
associated \susy breaking interactions which can give rise to neutrino
masses.  We evaluate the mass matrices for the various boson and
fermion modes and then discuss the effect that the lepton number
violating interactions have on the mass spectra using a
phenomenological precription to implement the experimental constraints
on the light neutrinos mass matrix.  We also discuss qualitatively the
conditions on the lepton number violating parameters set by the
unbounded from below directions, and from the absence of the 
charge and color breaking minima in this model.
\end{abstract}

\maketitle
                                                                                

\section{INTRODUCTION}
\label{sec1}

There are suggestive hints of lepton number violation in the
observation of neutrino flavor oscillations~\cite{Smirnov:2007pw}.
The link between the new physics beyond the Standard Model~(SM) and
lepton number violating interactions has so far been realized in
two main theoretical frameworks. The first option, the see-saw
mechanism~\cite{Minkowski:1977sc}, uses physics at the grand unified
scale, with lepton number violation represented by the
non-renormalizable interaction of dimension 5, ${\cal L}_{eff} =
\frac{1} {\Lambda} (LH_u)(LH_u)$, where $L$ and $H_u$ are the lepton
and Higgs doublets, respectively. The second option involves minimal
supersymmetric standard model~(MSSM) with broken~(generalized) lepton
parity, and uses bilinear and/or trilinear lepton number violating
couplings in the superpotential, with lepton number violation
occurring at the weak scale~\cite{Romao:2005jq}.

Recently, we have pointed out~\cite{Chemtob:2006ur} that in the
context of low energy supersymmetry there is an attractive option of
generating neutrino masses in the nonminimal supersymmetric standard
model~(NMSSM) with trilinear lepton number violation. Unlike the
minimal supersymmetric standard model~(MSSM), the superpotential of
the NMSSM is scale invariant. What is perhaps even more interesting is
the presence of an additional lepton number violating trilinear
coupling in the superpotential of NMSSM which has no analog in the
MSSM.  In the NMSSM with lepton number violation we have a situation
which is different from the one that arises in the see-saw mechanism
or the bilinear lepton number violation in MSSM in that no
dimensionful mass parameters~(large or small) are introduced.

There are several reasons for studying the nonminimal supersymmetric
standard model. First, and foremost, it is the simplest supersymmetric
extension of the Standard Model in which the electroweak scale
originates from the supersymmetry breaking scale only.  Second, and as
mentioned above, this model breaks lepton number and $R$-parity
explicitly through a unique trilinear superpotential
term~\cite{Pandita:1999jd, Pandita:2001cv}. Third, the model can
successfully explain the pattern of neutrino masses with normal type
of hierarchy~\cite{Chemtob:2006ur} without invoking either a large or
a small mass parameter.  And last, but not the least, its enlarged
Higgs sector may help in relaxing the fine-tuning and little hierarchy
problems of the MSSM~\cite{Dermisek:2005ar}, thereby opening new
perspectives for the Higgs boson searches at the next generation of
high energy colliders.

In addition to the familiar Yukawa interaction superpotential for the
three generations of quark and lepton superfields, and for the Higgs
bosons superfields, $ H_u,\ H_d$, the model couples the set of down
type Higgs boson and sleptons, $L_A = (H_d, L_i) \ ( A= d, i; \
i=1,2,3), $ with the Higgs singlet superfield $S $ through the
superpotential $ W= \tilde \l _A L_A H_u S - {\kappa\over 3} S^3 $,
where $ \tilde \l_A = (\l ,\ \ \tilde \l _i ) $ and $ \kappa $ are
dimensionless parameters.  The familiar soft \susy breaking trilinear
scalar couplings and mass terms introduce the dimensional parameters
$A_{\tilde \l _A}= (A_\l , A_{\tilde \l _i}),\ A_\kappa, $ and $ m^2
_{\tilde L , AB},\ m^2 _S $.  Whereas the fundamental parameters in
the action are $\tilde \l _A ,\ A_{\tilde \l _A} ,\ m^2 _{\tilde L ,
AB} $, the vacuum solution is specified by the dynamical parameters
corresponding to the vacuum expectation values~(VEVs) of the
electrically neutral scalar fields at the electroweak scale, $ v_u =
<H_u>,\ v_A = (v_d = < H_d>, v_i =<\tilde \nu _i>) $, which include
the sneutrinos together  with the neutral components of the Higgs
bosons. We use a covariant notation for the  down-type Higgs
boson and slepton superfields, and their couplings, in order to
illustrate the invariance of observables under the $SU(4)$ group of
transformations acting on the 4D column vectors $L_A$ and $\tilde \l
_A,\ A_{\tilde\l _A},$ and the matrix $ m^2 _{\tilde L , AB} $.
Strong correlations usually exist between the fundamental and the
dynamical parameters.  Thus, the experimental observations for the
light neutrino mass matrix tightly constrain certain combinations of
the parameters which characterize the inequivalent field directions
distinguishing the down-type Higgs boson from the sleptons, and hence
the interactions which conserve lepton number from those which violate
it. One may hope that imposing the necessary restrictions on the
misalignment type parameters would still leave some freedom in the
parameter space so as to allow the lepton number violating
interactions to significantly affect the physics of the NMSSM.  There
is a close analogy here with the situation that obtains in the MSSM with R-parity
violation~\cite{Grossman:1998py,gross99,haberbasis,grossman03,Davidson:2000uc,
Davidson:2000ne,davrius00,abada02}.

At the tree level, the size of the neutrino Majorana mass matrix is
knownn to be controlled by deviations from the $\tilde \l _A \propto
v_A$ alignment, such that the predicted single non-vanishing neutrino
mass is proportional to $\sum _i \tilde \l ^{'2}_i $, where the
effective misalignment parameters are defined by the linear
combinations $\tilde \l '_i \equiv \tilde \l _i - \l v_i /v_d $. We
have previously shown~\cite{Chemtob:2006ur} that the suitably
constrained NMSSM predicts an acceptable neutrino mass matrix of a
normal hierarchy type so that the tentative limits on the heaviest
neutrino mass, $ m_ {\nu _\tau } < (10 \ \text{MeV} \ - \ 1 \
\text{eV}) $, set bounds on the effective parameters given by $\tilde
\l'_i < ( 3.\ 10^{-2} \ - \ 10^{-5} ) $.  Bounds of similar type were
found for the couplings $\tilde \l _i $ in the basis $ v_i =0$ in an
attempt to fit an extended version of the model to the neutrino mass
differences and mixing angles~\cite{abada06}.  It has also been
observed that ratios as large as $\tilde \l _i /\l \sim 10^{-1} $
might be allowed at the price of fine tuning the NMSSM
parameters~\cite{Abada:2006qn}.  As far as one-loop contributions to
the neutrino mass matrix is concerned, the situation is more complex
due to the larger number of unknown parameters and the presence of
several sources of contributions arising in perturbation theory.  Thus,
to supress the contributions from the one-loop amplitude involving the
exchange of scalars and gauginos, it is necessary to impose strong
constraints on the effective parameters, $ B_i ' = B_i - B v_i / v_d
\ [B_A = (B, B_i) \equiv (A _{\tilde \l _i } \tilde \l _i , A_\l \l ) ]$ 
controlled by the $B _A \propto v_A$ alignment.

Supersymmetric models, including the nonminimal supersymmetric
standard model, contain a large number of scalar fields, such as the
scalar partners of quarks and leptons, which are not present in the
Standard Model.  As a result, the scalar potential of supersymmetric
models is fairly complicated. Such a potential can have minima which
could lead to the breakdown of color or electric charge conservation
through the vacuum expectation values of squark and slepton
fields. The condition that the global minimum of the theory does not
violate color or electromagnetic gauge invariance provides useful
constraints on the parameter space of the underlying supersymmetric
model, as has been well illustrated in the context of the minimal
supersymmetric standard model~\cite{Frere:1983ag,Claudson:1983et,
Nilles:1982mp, Komatsu:1988mt, Casas:1997ze} and its extended version
with broken R parity symmetry~\cite{Abel:1998ie, Hirsch:2004hr}.
Significant bounds can be obtained not only on the value of the
trilinear soft supersymmetry breaking parameters, but also on the
values of the bilinear soft scalar supersymmetry breaking parameter,
as well as scalar and gaugino masses, respectively.

The possibility that one can reproduce the lepton number violating 
observables  without imposing strong restrictions on the coupling
constants $\tilde \l _i$ motivates us to examine how the scalar sector
of the NMSSM is affected by treating $\tilde \l _i$ as free
parameters.  One may expect interesting implications regarding the
stability of the regular vacuum solution and the properties of the
neutral and charged scalars.  While performing the calculations within
a field basis independent
formalism~\cite{Grossman:1998py,gross99,haberbasis,
Davidson:2000uc,Davidson:2000ne,davrius00,abada02} is highly
desirable, this is very tedious and is not particularly
illuminating. It requires an unusually large amount of effort which is
not warranted at the present stage of our study. We, thus, choose to
pursue the calculations within a basis dependent formalism.  In
principle, there is a choice of basis for the neutrino superfields in
which the sneutrino VEVs can be set to zero at the expense of
transferring through a superfield redefinition the couplings $ L_i H_u
S$ to the trilinear lepton number violating couplings of the matter
superfields.  Although the calculations are greatly simplified in the
basis $ v_i=0$, there is a risk that this choice biases the parameter
space exploration, as has clearly been pointed out in the above
discussion regarding the neutrino mass matrix.  In fact, in the basis
choice $v_i =0$, the couplings $\tilde \l _i $ are so strongly
suppressed that the issue of establishing any observable effect on the
scalar sector would be closed from the very start.

In the present work, we shall adopt the view that the lepton number
violation phenomenology allows the couplings $\tilde \l _i $ to cover
a  wide range of variation, $ 0 \leq \vert \tilde \l _i / \l \vert
\leq 1$.  Thus, we treat all the $ \tilde \l _A $ on the same footing,
while requiring the soft and dynamical parameters to satisfy the near
alignment properties, $v _A \propto B _A $ and $v_A \propto \tilde \l
_A $, in agreement with the observations.  Following a
phenomenological approach, we treat $\tilde \l _i$ as free parameters,
and determine $v_i$ and $ B_i$ by restricting the linear combinations
$\tilde \l '_i = \tilde \l _i - \l v_i /v_d $ and $ B'_i = B_i - B v_i
/v_d$.  Since $ v_d, \ v_i$ are determined by minimizing the scalar
potential, this procedure implies a fine tuning.  However, one can use
symmetries to impose in a natural way a preferred direction in the
$L_A$ field space along which the various parameters get aligned.
While the horizontal flavor discrete
symmetries~\cite{banksgross95,borzum96} indeed have the ability to
enforce an approximate $\tilde \l _A \propto v_A$ alignment, this
property is robust only in the case of hierarchically small ratios,
${\tilde \l _i \over \l} \sim {v_i \over v_d} $, which is not exactly
what we want.  By contrast, the supergravity inspired models of the
type discussed by Nilles and Polonsky~\cite{nillespol97}, in which the
$\tilde \l _A L_A H_u S$ couplings arise from non-renormalizable
interactions controlled by a spontaneously broken $U(1)_R$  type
symmetry, are compatible with all the $\tilde \l _A$ having the same
size.  The alignment is realized   dynamically via universal type
boundary conditions on the soft parameters, and remains typically
stable under the renormalization group~\cite{nillespol97,hempfling97}.
Our main purpose is to study the impact of the lepton number violation
on the electroweak symmetry breaking vacuum of the nonminimal
supersymmetric standard model consistent with the experimental
measurements for the light neutrinos.  In the first stage, we discuss
its stability by testing for the occurrence of tachyons (negative
eigenvalues of the squared mass matrices) for the charged and neutral
scalar modes.  In the second stage, we analyze the dangerous
directions in the scalar field space along which the scalar potential
is unbounded from below~(UFB) or develops deeper charge and/or color
breaking~(CCB) minima.  Distinct methods must be used at these two
stages, which both involve the non-supersymmetric couplings in an
essential way.

This paper is organized as follows.  In Section~\ref{sec2} we describe
the nonminimal supersymmetric standard model with lepton number
violation, and set up our notation and conventions.  Complementary
definitions and results are summarized in the
Appendix~\ref{Notations}.  In Section~\ref{sec3} we discuss the
conditions on the superpotential couplings and supersymmetry breaking
couplings required to exclude the tachyon scalar modes for the regular
vacuum.  In Section~\ref{sec4} we explore the conditions on the
superpotential and supersymemtry breaking couplings implied by the
lifting of UFB directions, and the removal of CCB minima.  In
Section~\ref{sec5} we summarize our main conclusions.

\section{The NMSSM with Baryon and Lepton Number Violation} 
\label{sec2}

In this section we summarize the basic features of the superpotential
and the associated soft breaking terms for the nonminimal
supersymmetric standard model with baryon and lepton number violation,
and establish our notation and conventions.  Some of the details are
further discussed in the Appendix~\ref{Notations}.

\subsection{Superpotential and soft breaking terms} 

The superpotential of NMSSM with baryon and lepton number violation is
characterized by the scale invariant supepotential \bea W & = & W
_{RPC} +W _{RPV}, \label{nmssmrpv1} \eea where $W_{RPC}$ and $W_{RPV}$
are the baryon and lepton number~(and $R$-parity) conserving, and
baryon and lepton number~(and $R$-parity) violating contributions,
respectively to the superpotential~\cite{Chemtob:2006ur}.  These
contributions are written down explicitly in Appendix \ref{Notations}.
In the absence of lepton number conservation, there is no distinction
between the down-type Higgs~($H_d$) and the lepton superfields~($L_i$)
as they transform identically under the SM gauge group.  We can,
therefore, employ a four vector notation for the down type Higgs and
the lepton superfields, and the Yukawa couplings \bea L_A & =
& (H_d, L_i), \nonumber \\ \tilde \l _A & = & (\l, \tilde \l _i) , \
[A= d, \ i ; \ i=1,2,3],
\label{lepton1}
\eea which is given in detail in Eq.(\ref{yukawa1a}) of
Appendix~\ref{appexa}.  The quark and lepton generation indices are
denoted by the letters $i, j, \cdots $, and the trilinear couplings
obey the antisymemtry property, $\l _{ABk} = -\l_{BAk}.$ In this
notation we can write the complete superpotential of NMSSM with baryon
and lepton number violation as \bea W & = & \l ^u _{jk} H_u Q_j U_k^c
+ \ud \l _{ABk} L_AL_B E_k^c + \l '_{Ajk} L_A Q_j D_k ^c + \ud \l
''_{ijk} U^c _i D^c_j D^c _k \nonumber \\ & + & \tilde \l _A L_AH_u S
-{\kappa \over 3} S^3 .
\label{nmssmrpv2}
\eea The supersymmetric contribution to the Lagrangian of NMSSM with
baryon and lepton number violation generated by the superpotential
(\ref{nmssmrpv2}) can be obtained by a standard procedure.  To this we
must add the supersymmetric contribution from the $D$-terms generated
by the gauge interactions. In addition to the supersymmetric part, the
Lagrangian consists of soft supersymmetry breaking terms, which
include soft trilinear scalar couplings and soft masses for all
scalars and gauginos, respectively.  The part of the soft
supersymmetry breaking scalar potential of NMSSM containing trilinear
scalar couplings can be written as~(note that it is $-V$ that occurs
in the Lagrangian) \bea V^{\mbox{soft}}(\mbox{trilinear}) & = & V
_{RPC} ^ {\mbox{soft}}(\mbox{trilinear}) + V _{RPV} ^
{\mbox{soft}}(\mbox{trilinear}), \nonumber\\
& = & - \ud A ^\l _{ABk} \l _{ABk} \tilde L_A\tilde L_B \tilde E_k^c -
A ^{\l '} _{Ajk} \l '_{Ajk} \tilde L_A\tilde Q_j \tilde D_k ^c - A ^{u
} _{jk} \l ^u_{jk} H_u \tilde Q_j \tilde U^c _k \cr && - \ud A _{ijk}
^ {\l ''} \l ''_{ijk}\tilde U^c _i \tilde D^c_j\tilde D^c _k -
A_{\tilde \l _A} \tilde \l_A \tilde L_A H_u S - {A_{\kappa } \kappa
\over 3} S^3 + \ H.\ c., \eea and the part containing the bilinear
mass terms for the scalars and gauginos can be written as \bea
V^{\mbox{soft}}(\mbox{mass}) & = & V _{RPC}^{\mbox{soft}}(\mbox{mass})
+ V _{RPV} ^{\mbox{soft}}(\mbox{mass}), \eea with the lepton number
conserving RPC terms given by \bea V _{RPC}^{\mbox{soft}}(\mbox{mass})
& = & m^2 _{\tilde Q _i} \vert \tilde Q_i \vert ^2 + m^2 _{\tilde
U^c_i} \vert \tilde U^c_i \vert ^2 + m^2_{\tilde D^c_i} \vert \tilde
D^c_i \vert ^2
+ m^2 _{\tilde L_i } \vert \tilde L _i \vert ^2 + m^2 _{\tilde E^c_i}
\vert \tilde E^c_i \vert ^2, \nonumber\\ &+& m^2 _{H_u } \vert H_u
\vert ^2 + m^2 _{H_d } \vert H_d \vert ^2 + m^2 _{S } \vert S \vert ^2
\nonumber\\ &-& ( \sum _{a=3,2,1} \ud M_a \tilde \l _a \tilde \l _a +
\ H. \ c.  ) , \eea and the lepton number violating terms given by
\bea V _{RPV}^{\mbox{soft}}(\mbox{mass}) & = & m^2_{H_d \tilde L_i }
H_d ^\dagger \tilde L_i + \mu _{H_d H_u } ^2 H_d H_u + \mu _{\tilde
L_i H_u} ^2 \tilde L_i H_u + \ H.c.  \eea We have used the convention
in which the repeated indices are implicitly summed over and have
suppressed the $SU(3)$ color quantum numbers of the quark and squark
fields by setting, for instance, $ \tilde Q_i ^\a = (\tilde U_i ^\a ,
\tilde D_i ^\a ) \to \tilde Q_i = (\tilde U_i , \tilde D_i ) $.
Although we have represented the fields mass mixing by the general
terms, $m^2 _{\tilde L , ij } \tilde L _i ^\dagger \tilde L _j$ and
$m^2 _{\tilde L , AB } \tilde L _A ^\dagger \tilde L _B$,  for the
sleptons and for the down Higgs boson and sleptons, we neglect
thoughout the present work the intergenerational mixing of sleptons by
assuming that $\tilde L_i$ are mass basis fields with $m^2 _{\tilde L
, ij } = m^2 _{\tilde L_i } \d _{ij}$ while retaining the
off-diagonal mass parameters, $ m^2_{H_d \tilde L_i } \equiv m^2_{di} \ne 0$.  Since \susy breaking respects the SM gauge
symmetry, we must assign the same soft mass parameters for the two
members of electroweak doublet fields, as is explicitly done for the
expression of the slepton mass terms, viz.  $ m_{\tilde L_i } ^2 \vert
\tilde L_i \vert ^2 = m_{\tilde L_i } ^2 (\vert \tilde \nu _i \vert ^2
+ \vert \tilde e_i \vert ^2 )$.  The choice of the sign of $M_2$ in
the chargino mass matrix conforms with the convention most often used
in the literature.  We have also included above, for generality, the
holomorphic mass mixing terms with the parameters $ \mu _{Au} ^2 =
(\mu _{H_d H_u } ^2 \equiv \mu _{du} ^2 , \ \mu _{\tilde L_i H_u} ^2
\equiv \mu _{iu} ^2) $, although the soft bilinear operators $\mu
_{Au} ^2 \tilde L_A H_u $ are, in principle, disallowed, since the
same discrete symmetry that forbids the bilinear superpotential terms
$ \mu _A L _A H_u$ should also forbid the  bilinear mass mixing operators.  
Mass mixing terms of same structure would still arise as effective 
contributions through the soft trilinear operators, 
$ - A _{\tilde \l_A } \tilde \l_A <S> \tilde L_A H_u$, 
with $ \mu _{Au} ^2 = - A _{\tilde \l_A }\tilde \l_A <S>.$

The scalar potential receives contributions from the supersymmetric as
well as the soft supersymmetry breaking interactions involving the
electrically neutral and charged complex scalar fields. The
supersymmetric contribution arises from the superpotential
(\ref{nmssmrpv2})~(the $F$-terms), and from the $D-$terms.  The Higgs
boson and slepton $F$-term contributions to the scalar potential can
be written as \bea V_F(\mbox{Higgs, sleptons}) & = & \vert W _{H_u ^0}
\vert ^2 + \vert W _{H_u ^+} \vert ^2 + \vert W _{S } \vert ^2 + \vert
W _{\tilde \nu _A} \vert ^2 + \vert W _{\tilde e _A} \vert ^2 + \vert
W_{\tilde e_i^c} \vert^2,
\label{fslh1}
\eea where \bea W _{H_u ^0} & = & -\l ^u_{jk} u_j u_k ^c +\tilde \l _A
v_A x,\, \, \ W _{H_u ^+} =\l ^u_{jk} d_j u_k ^c -\tilde \l _A e_A x,
\label{fh1}  \\ 
W _S & = & \tilde \l _A v_Av_u - \tilde \l _A e_A v_+ -\kappa x^2 ,
\label{fh2}\\
W _{\tilde \nu _A } & = & \l_{ABk} e_B e_k^c + \l '_{Ajk} d_j d_k ^c +
\tilde \l _A v_u x, \label{fsl1} \\ W_{\tilde e_A } & = & - \l '_{Ajk}
u_j d_k ^c - \l _{BAk} v_B e_k ^c -\tilde \l _A v_+ x, \label{fsl2} \\
W_{\tilde e_i^c} & = & \l_{ABi} v_A e_B, \label{fsl3} \eea with
various vacuum expectation values denoted by \bea v_A & = & (v_d =
<H_d^0>, v_i = <\tilde \nu _i > ), \, \, e_A = (v_{-} = <H_d^->, e_i
=<\tilde e_i > ), \label{vev1} \\ e_i^c & = & <\tilde e_i^c>, \, \,
v_+= <H_u^+>,\, \, v_u = <H_u^0>.
\label{vev2} 
\eea The $F$-term contribution of the squarks to the scalar potential
can be written as \bea V_F(\mbox{squarks}) & = & \vert W _{u_i} \vert
^2 + \vert W _{d_i } \vert ^2 + \vert W _{u^c_i } \vert ^2 + \vert W
_{d^c _i } \vert ^2,
\label{fsq}
\eea where \bea W _{u_i } & = & -\l ^u_{ik} v_u u_k ^c - \l '_{Aik}
e_A d_k^c, \, \, \, \, \, \, W _{d_i } = \l ^u_{ik} v_+ u_k ^c + \l
'_{Aik} v_A d_k^c, \label{fsq1} \\ W _{u_i ^c} & = & \l ^u_{ji} ( v_+
d_j - v_u u_j) +\ud \l '' _{ijk} d_j^c d_k^c, \, \, W _{d_i^c } =
\l'_{Aji} (v_A d_j -e_A u_j), \label{fsq2}
\eea with the vacuum expectation values of the squarks defined by \bea
u_i & = & <\tilde U _i>,\ d_i = <\tilde D _i >,\ u_i ^c = <\tilde U _i
^c> , \ d_i ^c = <\tilde D_i ^c>. \label{vev3} \eea The $D$-term
contributions to the scalar potential can be written as \bea V_D & = &
V_D^{U(1)} + V_D^{SU(2)} + V_D^{SU(3)}, \label{vd} \eea with \bea
V_D^{U(1)} & = & {g_1^2 \over 8} \left(-\vert v_d\vert ^2 -\vert v_-
\vert ^2 + \vert v_u \vert ^2 +\vert v_+\vert ^2 + {1\over 3}\vert
u_i\vert ^2 + {1\over 3} \vert d _i\vert ^ 2 \right. \nonumber \\ & &
- \left. {4 \over 3} \vert u _i^{c} \vert ^2 + {2\over 3} \vert d
_i^{c} \vert^2 - \vert v_i \vert^2 - \vert e_i \vert ^2 + 2 \vert e ^c
_i \vert ^2 \right)^2, \label{vd1}\\ V_D^{SU(2)} & = & {g_2^2 \over
8}\left[ \left( \vert v_d\vert ^2 - \vert v_- \vert^2 + \vert v_+
\vert ^2 - \vert v_u \vert ^2 +\vert v_i \vert ^2 - \vert e_i \vert ^2
+ \vert u _i\vert ^2 - \vert d _i\vert ^2 \right)^2 \right.  \nonumber
\\ & & \left . + 4 \vert v_d ^\star v_ - + v_i ^\star e_i + v_+ ^\star
v_u + u_i^\star d_i \vert ^2\right], \\ V_D^{SU(3)} & = & {g_3^2 \over
6} \bigg (\vert u _i \vert ^2 + \vert d_i \vert ^2 - \vert u^c_i \vert
^2 - \vert d ^c_i \vert ^2 \bigg )^2.  \eea

The complete soft \susy breaking scalar potential is given by \bea
V^{soft} & = & m^2 _{\tilde Q _i} ( \vert u_i \vert ^2 + \vert d_i
\vert ^2) + m^2 _{\tilde U^c_i} \vert u^c_i \vert ^2 + m^2 _{\tilde
D^c_i} \vert d^c_i \vert ^2 + m^2 _{\tilde L_i } (\vert v_i \vert ^2 +
\vert e_i \vert ^2 ) \nonumber \\ & & + m^2 _{\tilde E^c_i} \vert
e^c_i \vert ^2 + m^2 _{H_d } (\vert v_d \vert ^2 + \vert v_- \vert ^2
) + m^2 _{H_u } ( \vert v_u \vert ^2 + \vert v_+ \vert ^2)+ m^2 _{S }
\vert x \vert ^2 \nonumber \\ & & + \bigg [\mu ^2 _{Au} (v_A v_u - e_A
v_+) + m ^2 _{H_d \tilde L_i} (v_d v_i^\star + v_- e_i^\star ) + H.\
c.\ \bigg ] \nonumber \\ & & + \left [- A _{\tilde \l _A } \tilde \l
_{A} ( v_A v_u - e_A v_+ ) x - {1 \over 3 } A_\kappa \kappa x ^3 + A ^
u _{jk} \l ^u _{jk} ( u_j v_u -d_j v_+)u_k^c - A ^\l _{ABk} \l _{ABk}
v_A e_B e_k ^c \right . \nonumber \\ & & \left . - A ^{\l '} _{Ajk} \l
'_{Ajk} ( v_A d_j -e_A u_j) d_k ^c - \ud A _{ijk} ^ {\l ''} \l
''_{ijk} u^c _i d^c_j d^c _k + H.\ c.\ \right ] .  \eea We recall our
convention of summing over the repeated indices for the (suppressed)
color indices and for the squark and slepton generation labels, $i,\ j
, \ k ,\ \cdots= 1, 2 , 3 $.  Assuming that the squark fields point in
a fixed direction in the color space, no summations over the color
indices would be present in the above formulae.  Furthermore, assuming 
that the VEVs for the squarks and sleptons of different generations affect the
scalar potential independently of each other, allows one to disregard
the summations over the generation labels.

In order to discuss NMSSM in a somewhat realistic manner, it is
necessary to include the one-loop contributions to the scalar
potential.  We shall make the usual
assumption~\cite{Pandita:1993hx,Pandita:1993tg} that the scalar
potential is dominated by the top quark and squark modes whose
explicit contributions are given by \bea V_{loop} & = & {3 \over 32
\pi ^2 } \bigg [ \sum _{i=1,2} m^4 _{\tilde t_i } (\ln { m^2_{\tilde
t_i } \over Q ^2 } - {3 \over 2} ) -2 \bar m^4 _{t } (\ln {\bar m^2_{t
} \over Q ^2 } - {3 \over 2} ) \bigg ] , \label{eqcolwei} \eea where
$\bar m_t = \l _t v_u $ is the top quark mass and $ m^2_{\tilde t_i }
,\ (i=1,2)$ denote the squared masses of the top squarks.  We do not
discuss this point any further, as detailed discussions can be found
in Refs.~\cite{Pandita:1993hx, Pandita:1993tg, miller04,
Barger:2006dh}.

\subsection{Symmetry Constraints on the Parameter Space  and  Choice 
of Free Parameters}

The scalar sector involves the scalar field components of the doublet
and singlet Higgs boson superfields, and the lepton and quark superfields.  
The parameter space of the NMSSM with lepton number violation consist of
the gauge couplings $g_a~(a=3, 2, 1)$, the \susyq couplings $ \l, \
\kappa,\ \l ^u _{jk} ,\ \l ^d _{jk} ,\ \l ^e _{jk} ,\ \tilde \l_i ,\
\l _{ijk},\ \l '_{ijk}, $ the soft \susy breaking couplings $ A_
{\tilde \l _A} \tilde \l _A , \ A_\kappa \kappa,\ A ^\l _{ABk} \l
_{ABk},\ A^{\l '} _{Ajk} {\l '} _{Ajk} ,\ A^u _{jk} \l ^u _{jk} $, and
the soft \susy breaking mass parameters, $ m^2 _{\tilde Q _i}, \ m^2
_{\tilde U^c _i}, \ m^2 _{\tilde D^c _i} ,\ m^2_{H_d},\ m^2_{H_u}, \
m^2_{\tilde L_i }, m_S^2, \ m^2_{\tilde E^c_i }, \ m^2 _{ H_d\tilde
L_i} ,\ \mu ^2 _{Au}.$ We summarize below some useful definitions and
abbreviations used in this paper: \bea && m_W ^2 = {g_2 ^2 \over 2}
(v_u ^2 + \hat v_d^2 ), \ m_Z ^2 = { g_1^2 + g_2 ^2 \over 2} (v_u ^2 +
v_d^2 + v_i^2 ), \cr && \tan \b = {v_u\over v_d} , \ G^2 _{\pm } =
{g_1 ^2 \pm g_2 ^2 \over 8}, \ \ v^2 = v_u ^2 + \hat v_d^2, \ \ \hat
v_d^2 = v_d^2 + v_i^2 .\eea The electroweak symmetry breaking scale
has a numerical value $ v = {\sqrt 2 m_W \over g_2 } \simeq 174 $ GeV.

Without loss of generality, the parameterization of the NMSSM can be
simplified by using the symmetries of the action and the independence
of observables under phase redefinitions of the fields.  The
invariance of the Lagrangian under the $SU(2)_L \times U(1)_Y$ gauge
symmetry allows eliminating four real field degrees of freedom,
independently of the structure of the Yukawa couplings.  We choose the
convention where the VEVs of the up-type Higgs boson electroweak
doublet are set as $ v_+ \equiv <H_u^+> = 0 $ and $ v_u \equiv <H_u^0>
\in R ^+ $.  With the choice $ v_+=0$, the minimization with respect
to the field $ H_u ^+$ becomes trivial.  Next, using the scalar
potential invariance under phase redefinitions of the fields $S,\ H_d,
\ \tilde U^c , \tilde Q $ and the pair of fields $ H_d ^-, \ \tilde
D^c $, one can make the following choice~\cite{Ellis:1988er} for
various parameters \bea && \kappa A_\kappa \in R ^+,\ \l A_{\l } \in R
^+ ,\ A ^u \l _u \in R ^+,\ u _i \in R^+ ,\ d_i \in R^+.  \eea For
completeness, we also observe that in the presence of lepton number
violation the NMSSM still satisfies a Peccei-Quinn symmetry in the
limit $\kappa \to 0 $ and a R type symmetry in the limit $ A_\l \to 0
,\ A_{\tilde \l _i} \to 0 $, where the $ U(1)_{PQ} $ and $ U(1)_{R} $
groups are defined by the assignment of charges for $(Q, U^c, D^c, L,
E^c, H_d, H_u, S), \quad Q_{PQ}= (-1,0,2,-1,2,-1,1,0),\ Q_{R}= {1\over
3} (3,0,2,1,4,1,3,2). $ In the limit $x \to \infty$, with $\tilde \l
_A x = -\mu _A ,\ \kappa x ,\ A_ {\tilde \l _A} \tilde \l _A x = - B_A
\mu _A $ fixed, the physical observables must reduce to those of the
MSSM with bilinear $R$-parity violation~\cite{Hirsch:2004hr}.

In order to obtain significant contributions to the scalar sector
observables from the lepton number violating interactions, some subset
of the parameters $\tilde \l _i ,\ A_ {\tilde \l _i } , \ m ^2 _{H_d
\tilde L_i} , \ v_i $ should assume large enough values.  Whether this
can be achieved while imposing at the same time highly suppressed
contributions to the neutrino mass matrix is closely related to the
formal symmetry under the $SU(4)$ group symmetry in the field space
$L_A = (H_d , L_i )$.  The freedom with respect to the choice of $L_A$
field basis entails that physical observables can only depend on the
$SU(4)$ invariant combinations of the parameters, $\tilde \l _A , \ B
_A = A _{\tilde \l _A } \tilde \l _A ,\ m^2 _{\tilde L , AB} ,\ v_A$,
transforming as vectors or tensors. The physical observables can only
depend on the singlet combinations.  At the quantum level, the
field basis independence holds only after summing all contributions 
at a given order of the  perturbation theory.  The basis independent formalism
can be developed along similar lines as for the MSSM with R-parity
violation~\cite{Grossman:1998py,gross99,haberbasis}.  Since $ H_d$ is
distinguished from $L_i$ by not having  lepton number, it follows
that the lepton number violating observables can only depend on the
invariants of the angle type characterizing the inequivalent
directions assigned to $H_d$.  Thus, the tree level contributions to
the neutrino mass matrix involve only the misalignment parameter
$\vert \tilde \l \wedge v \vert ^2 = \ud \sum _{A,B} ( \tilde \l
\wedge v ) ^2 _ {AB} \equiv \ud \sum _{A,B} ( \tilde \l _A v_ {B} -
\tilde \l _B v_ {A} ) ^2,$ while the one-loop contributions from the
scalar-neutralino exchange Feynman diagrams, for instance, involve the
invariant misalignment parameters, $\vert B \wedge v \vert $ and $
\vert B \wedge \tilde \l \vert $.  By contrast, the lepton number
conserving observables can also depend on the scalar products $ \hat
v_d ^2 = \sum _A v_A ^2,\ v\cdot B = \sum _A v_A B_A, $ and $\tilde \l
^T m ^ 2 _{\tilde L } v = \tilde \l _A m ^ 2 _{\tilde L , AB} v _B $.

To avoid the excessive effort involved in developing the basis
independent formalism, we have pursued the analysis by making a fixed
choice of the $L_A$ field basis, while distinguishing between the free
and constrained parameters phenomenologically.  Although the choice $
v_i =0$ is perfectly admissible, this biases the exploration of the
parameter space.  Upon working in the basis choice $ v_i \ne 0$, the
constraints from the tree and loop level contributions to the neutrino
mass matrix can be implemented in several ways.  Since the tree
contribution is proportional to $\tilde \l ' _i \equiv \tilde \l _i -
\l v_i / v_d $, we choose  to regard $\tilde \l _i $ as free
parameters and assign by hand $v_i $ consistently with the restricted
range of variation for $\tilde \l ' _i $.  The basic relation between
the gauge boson mass and the electroweak breaking mass scale, $ v ^2
\equiv v ^2 _u+ v ^2 _d+ v_i ^2 =v ^2 _u+ \hat v ^2 _d $, is then
implemented by using the parameterization of the Higgs-sneutrinos VEVs
\bea && v_d ={v_u \over \tan \b } = {v \over (1+\s _i ^2 +\tan ^2 \b )
^\ud } ,\quad v_i = \s _i v_d ,\qquad [\s _i = {\tilde \l _i -\tilde
\l ' _i \over \l } ],
\label{eqmatchscale} 
\eea where we retain the usual definition for the ratio of VEVs, $\tan
\b = v_u /v_d $.  At the one-loop level, the contributions to the
neutrino mass matrix from the neutralino-slepton exchange Feynman
diagram with a double mass insertion, as given by Fig. $1(B)$ and
Eq.~(III.54) of our previous work~\cite{Chemtob:2006ur}, bounds the
misalignment parameter $ \vert B\wedge v \vert $, or equivalently the
effective parameters $ \eta _i\equiv {B_i\over B} - {v_i\over v_d} =
{A_{\tilde \l _i } \tilde \l _i \over A_\l \l } - {v_i\over v_d} $.
With the restrictions from the neutrino masses set on $\tilde \l '_i $
and $ \eta _i $, the parameters $v_i $ and $A_{\tilde \l _i } $ are
explicitly determined in terms of $\tilde \l _i $.  It is important to
note that once the equations of motion are satisfied, the conditions $
\tilde \l _A \propto v _A $ and $ B_A \equiv A_{\tilde \l _A } \tilde
\l _A \propto v _A$ entail the condition on the squared mass matrix $
m^2_{\tilde L, AB} v_B \propto v _A $.  It then follows that fixing
$\tilde \l '_i$ and $ \eta _i $ still leaves the freedom of choosing
the slepton mass parameters $ m^2 _{\tilde L _i } $.  Because of the
strongly suppressed values $\tilde \l ' _i << 1$ and $\eta _i << 1$
imposed by the neutrino masses, the precise values assigned to $\tilde
\l ' _i $ and $\eta _i $ have little effect on the final results.

We shall develop the study of the regular vacuum solution of the
scalar sector in terms of the neutral scalar field VEVs $ v_d, \
v_u,\ x $ and $ v_i$.  A necessary condition for stability is obtained
by testing for the absence of saddle points of the scalar potential
along the charged and neutral boson field directions in the field
space.  This condition is equivalent to the requirement  that the 
squared mass matrices for the charged and neutral scalar bosons 
are free from negative squared mass tachyonic eigenvalues.  
To simplify the discussion we shall assign
finite values for the lepton number violating parameters (coupling
constants and VEVs) one at a time for each lepton number flavor, so
that finite coupling constants $\tilde \l _i ,\ A _{\tilde \l _i } ,
\ m^2 _{H_d \tilde L_i} $ and finite VEVs, $v_i $, are assigned for a
fixed flavor index $i$.  We thus neglect the quadratic terms of the
form $ \tilde \l _i \tilde \l _j, \ v_i v_j,\ \tilde \l _i v_j, \
\cdot \cdot \cdot $, with $ i\ne j $.  The single dominant coupling
constant assumption is not too restrictive as long as the lepton
number violating contributions are small relative to those of the
lepton number conserving contributions.  To assess the validity of
this approximation, one could envision using the variant prescription
selecting some linear combinations of parameters in the $SU(4)$ group
of the $L_A $ fields.

The structure of the scalar mass matrices depends on the way in which
one implements the equations of motion.  Since the scalar potential of
the NMSSM is given by a polynomial in the VEVs of the neutral fields,
it is impossible, in general, to obtain analytic formulae for the said
VEVs.  Instead, we follow the practical procedure in which one
eliminates the soft mass parameters $ m^2 _{H_u }, \ m^2 _{H_d}, \ m^2
_{S}$ via the minimization conditions for the VEVs of electrically
neutral fields $ v_u,\ v_d ,$ and $ x $, respectively.  For the
sneutrino VEVs, $v_i$, however, we need to specify beforehand our
choice for the independent mass parameters, because of the presence of
the off-diagonal and diagonal type sleptons mass parameters for the
combination $ (H_d, \ \tilde L_i )$, namely $m^2_{ H_d \tilde L_i}$
and $m_{\tilde L , ij} ^2 $.  The freedom in solving the field
equations (or equations of motion) for the sneutrinos $\tilde \nu _i$
could be used to eliminate the mass parameters $m^2_{ H_d \tilde
L_i}$, which would then leave us with $ m ^2 _{\tilde L_i } $ as free
parameters.  We call this option in the following as our prescription
I.  Alternatively, we could eliminate the mass parameters $m ^2
_{\tilde L_i } $, and hence use $m ^2_{ H_d \tilde L_i}$ as free
parameters.  We call this option in the following as prescription II.
These prescriptions only differ in the way one treats the input data
for the soft masses.  While prescription II is perhaps more natural,
since $m ^2_{ H_d \tilde L_i}$ are lepton number violating parameters
on the same footing as the soft parameters $A_{\tilde \l _i}$, this
has the drawback of introducing inverse powers of the parameters $
v_i$, which are expected to be small.  In any case the results in
prescription II are readily obtained from those in prescription I by
substituting the expression for $ m ^2 _{\tilde L_i } $ implied by the
equations of motion for $\tilde \nu _i $, which we explicitly provide
in Eq.~(\ref{eomneut}) of Appendix~\ref{appexa}.  For definiteness, we
quote below the formulae obtained with prescription I.

\subsection{Mass Matrices of  Scalar Bosons}  
   
The lepton number violating term proportional to $\tilde\lambda_i$ in
the superpotential (\ref{nmssm2}) produces mixing between the charged
Higgs bosons and scalar leptons~\cite{Chemtob:2006ur}.  Similarly,
there will be mixing between the charginos and the charged
leptons. These mixings can be studied through the appropriate mass
matrices. Here we discuss the mass matrices of charged Higgs
bosons/charged scalar leptons, and neutral Higgs bosons/ neutral
scalar leptons, respectively.  The mass matrices for charginos/charged
leptons and neutralinos/neutral leptons are discussed in the next
subsection. The field basis for the charged Higgs-slepton scalar modes
is denoted by the column vectors \bea && \Phi _{ch} = (H_d ^{-\star }
,\ H_u ^+ ,\ \tilde e_i^\star ,\ \tilde e_i ^c ),\ \Phi ^\dagger _
{ch} = (H_d ^{- } ,\ H_u ^ {+ \star } ,\ \tilde e_i,\ \tilde e_i ^ {c
\star } ), \eea with the mass term in the Lagrangian given by \bea
-L_{mass} = \Phi_{ch}^\dagger M^2 _{ch} \Phi _{ch} + \ H.\ c.  \eea
After a straightforward calculation, we obtain the mass squared matrix
for the charged scalars whose elements can be written as \bea M^2
_{H_d H_d ^\star } &=& {{v_u} \over 2 v_d} ( -2 \mu ^2_{du} + {g_2^2}
{v_d} {v_u} - 2 {{\l }}^2 {v_d} {v_u} + 2 {A _\l \l } {x} + 2 {\kappa
} {\l } {{x}}^2 ) - {v_u {x} {v_i} \tilde \l_i \over v_d ^2 } (
{A_{\tilde \l _i} }+ {\kappa } {x} ) + {v_u v_i \mu ^2_{iu} \over v_d
^2 } \nonumber \\ & + &{ v_i ^2 \over 4 v_d ^2} ( 4 {m^2 _{\tilde L_i}
} + {g_1^2} {{v_d}}^2 - {g_2^2} {{v_d}}^2 - {g_1^2} {{v_u}}^2 -
{g_2^2} {{v_u}}^2 + 4 {{\tilde \l _i }}^2 {{v_u}}^2 + 4 {{\tilde \l _i
}}^2 {{x}}^2 ) \nonumber \\ & + & { v_i^4 \over 4 v_d^2} ( {g_1^2} +
{g_2^2} ) +\l _{dji} \l _{dki} v_j v_k , \\ M^2 _{H_d H_u } &=&
\frac{{g_2^2} {v_d} {v_u}}{2} - {{\l }}^2 {v_d} {v_u} + \l {x} ({A _\l
} + {\kappa } x ) - {\l } {\tilde \l _i } {v_u} {v_i} - \mu ^2 _{du },
\\ M^2 _{H_d \tilde e_i ^\star } & = & {\l } {\tilde \l _i } {{x}}^2 +
\frac{{g_2^2} {v_d} {v_i}}{2} - \l _{dAj} \l _{Bij} v_A v_B , \\ M^2
_{H_d \tilde e^c_i } & = & - \l _{dAi} \tilde \l _A v_u x + A ^\l
_{dAi} \l _{dAi} v_A , \\ M^2 _{H_u ^\star H_u } & = & {v_d\over 2v _u
} (-2 \mu ^ 2_{du} + {g_2^2} {v_d} {v_u} - 2 {{\l }}^2 {v_d} {v_u} + 2
{A _\l \l } {x} + 2 {\kappa } {\l } {{x}}^2 ) \nonumber \\ & + &{ v_i
\over v _u } ( -\mu ^2_{iu} -2 {\l } {\tilde \l _i } {v_d} {v_u} + {x}
\tilde \l_i ( {A_{\tilde \l _i} } + {\kappa } {x} ) ) + \frac{(
{g_2^2} - 2 {{\tilde \l _i }}^2 ) {{v_i}}^2}{2}, \\ M^2 _{H_u ^\star
\tilde e_i ^\star } &=& -\mu ^ 2 _{iu} - {\l } {\tilde \l _i } {v_d}
{v_u} + {x} \tilde \l_i ( {A_{\tilde \l _i} } + {\kappa } {x} ) + {
v_u v_i \over 2 } ({g_2^2} - 2 {{\tilde \l _i }}^2 ), \\ M^2 _{H_u
^\star \tilde e_i ^ c } & = & \l _{BAi} \tilde \l _A x v_B , \\ M^2 _{
\tilde e_i \tilde e_j ^\star } &=& {m^2 _{\tilde L, ij } } + {1 \over
4 } [ (g_1^2-g_2^2) ( {{v_d}}^2 - {{v_u}}^2 ) + ( {g_1^2} + {g_2^2}
){{v_i}}^2 + 4 {{\tilde \l _i }}^2 {{x}}^2] \d _{ij} + \l _{Aik} \l
_{Bjk} v_A v_B , \\ M^2 _{\tilde e_i \tilde e_j ^ c } & = & \tilde \l
_A v_u \l _{Aij} x - A ^\l _{Aij} \l _{Aij} v_A , \\ M^2 _{\tilde e_i
^ {c\star } \tilde e_j ^ c } & = & m^2_{\tilde E^c , ij } + \ud
{g_1^2} ( -{{v_d}}^2 - v_i^2 + {{v_u}}^2 ) \d _{ij} + \l _{ABi} \l
_{AB'j} v_B v _{B'} .  \eea For completeness, we have quoted the
results for general bases of the slepton fields. The above formulae
agree with Ref.~\cite{hambye00} for the MSSM with R parity violation.

We now consider the neutral spin-$0$ Higgs-sneutrino fields whose real
and imaginary parts carry positive and negative CP quantum numbers.
The field basis for the CP-even and CP-odd Higgs-sneutrino scalar
modes are defined by the Lagrangian mass terms as \bea -L _{mass} & =
& \Phi _{neut}^\dagger M^2 _{neut} \Phi _{neut} + \ H.\ c. \nonumber
\\ & = & \Phi _{s,i }^\dagger M^2 _{s, ij } \Phi _{s, j } + \Phi _{p,i
}^\dagger M^2 _{p, ij } \Phi _{p, j }, \eea where the neutral spin-$0$
fields have the decomposition \bea \Phi _{neut ,i } & = & {1\over
\sqrt 2 } (\Phi _{s , i } + i \Phi _{p , i} ), \ \ \Phi_{neut ,i } = (
H_d ^0, \ H_u ^0,\ S,\ \tilde \nu _j ).  \eea The mass squared matrix
for CP-even scalars in the prescription I, using the equations of
motion for $\tilde \nu _i$ to eliminate the mass parameter $m ^2_{ H_d
\tilde L_i}$ with $ m ^2 _{\tilde L _i} $ as free parameters, is given
by \bea M^2 _{s, dd} & = &\frac { 1 } {{v_d}} [ 4 {G_+^2} {{v_d}}^3 +
{v_u} {x} ( {A _\l \l } + {\kappa } {\l } {x} ) ] - \frac{{v_u}
{x}{v_i} } {{{v_d}}^2} ( {A _{\tilde \l _i} \tilde \l _i} + {\kappa }
{\tilde \l _i} {x} ) \nonumber \\ && + \frac{{v_i}^2} {{{v_d}}^2} (
{m^2 _{\tilde L _i }} + 2 {G_+^2} ( {{v_d}}^2 - {{v_u}}^2 ) + {{\tilde
\l _i}}^2 ( {{v_u}}^2 + {{x}}^2 ) ) + \frac{2 {G_+^2} {{v_i}}^4}
{{{v_d}}^2}, \\ M^2 _{s, du } & = & -4 {G_+^2} {v_d} {v_u} + 2 {{\l
}}^2 {v_d} {v_u} - {A _\l \l } {x} - {\kappa } {\l } {{x}}^2 + 2 {\l }
{\tilde \l _i} {v_u} {v_i},\\ M^2 _{s, dS } & = &- {A _\l \l } {v_u} +
2 {\l } ( {\l } {v_d} - {\kappa } {v_u} ) {x} + 2 {\l } {\tilde \l _i}
{x} {v_i},\\ M^2 _{s, d \tilde \nu _i} & = & \frac{ 1} {v_d} [{v_u}
{x} ( {A _{\tilde \l _i} \tilde \l _i} + {\kappa } {\tilde \l _i} {x}
) - ( {m^2 _{\tilde L _i }} - 2 {G_+^2} ( {{v_d}}^2 + {{v_u}}^2 ) +
{{\tilde \l _i}}^2 ( {{v_u}}^2 + {{x}}^2 ) ) {v_i} \nonumber \\ && - 2
{G_+^2} {{v_i}}^3 ], \\ M^2 _{s, uu} & = & \frac{1} {v_u} [ 4 {G_+^2}
{{v_u}}^3 + {v_d} {x} ( {A _\l \l } + {\kappa } {\l } {x} ) + {x} ( {A
_{\tilde \l _i} \tilde \l _i} + {\kappa } {\tilde \l _i} {x} ) {v_i} ]
,\\ M^2 _{s, uS} & = & - {A _\l \l } {v_d} + 2 ( -( {\kappa } {\l }
{v_d} ) + ( {{\l }}^2 + {{\tilde\l _i}}^2 ) {v_u} ) {x} + ( -{A
_{\tilde \l _i} \tilde \l _i} - 2{\kappa } {\tilde \l _i} {x} )
{v_i},\\ M^2 _{s, u\tilde \nu _i } & = & 2 {\l } {\tilde \l _i} {v_d}
{v_u} - {x} ( {A _{\tilde \l _i} \tilde\l _i} + {\kappa } {\tilde \l
_i} {x} ) + 2 ( -2 {G_+^2} + {{\tilde \l_i}}^2 ){v_u} {v_i} ,\\ M^2
_{s, S S} & = & \frac{{A _\l \l } {v_d}{v_u}}{{x}} + {x} ( -{A _\kappa
\kappa } + 4 {{\kappa }}^2 {x} ) +\frac{{A _{\tilde \l _i} \tilde \l
_i} {v_u} {v_i}}{{x}},\\ M^2_{s, S\tilde \nu _i } &=& - {A _{\tilde \l
_i} \tilde \l _i} {v_u} + 2{\l } {\tilde \l _i} {v_d} {x} - 2 {\kappa
} {\tilde \l _i} {v_u} {x} + 2 {{\tilde \l _i}}^2 {x} {v_i} , \\ M^2
_{s,\tilde \nu _i \tilde\nu _i } &=& {m^2 _{\tilde L _i }} + 2 {G_+^2}
( {{v_d}}^2 - {{v_u}}^2 ) + {{\tilde \l _i}}^2 ( {{v_u}}^2 + {{x}}^2 )
+ 6 {G_+^2} {{v_i}}^2 , \eea and for the CP-odd modes by \bea M^2 _{p,
dd} & = &\frac{{v_u} {x} ( {A _\l \l } + {\kappa } {\l } {x} ) }{
{v_d}} - \frac{{v_u} {x} ( {A _{\tilde \l _i} \tilde \l _i} + {\kappa
} {\tilde \l _i} {x} ) {v_i}}{{{v_d}}^2} \nonumber \\ && + \frac{(
{m^2 _{\tilde L _i }} + 2 {G_+^2} ( {{v_d}}^2 - {{v_u}}^2 ) + {{\tilde
\l _i}}^2 ( {{v_u}}^2 + {{x}}^2 ) ) {{v_i}}^2}{{ {v_d}}^2} + \frac{2
{G_+^2} {{v_i}}^4} {{{v_d}}^2}, \\ M^2 _{p, du } & = & {x} ( {A _\l \l
} + {\kappa } {\l} {x} ), \\ M^2 _{p, dS } & = & {v_u} ( {A _\l \l } -
2 {\kappa } {\l} {x} ), \\ M^2 _{p, d \tilde \nu _i} & = & \frac{{v_u}
{x}({A_{\tilde \l _i}\tilde \l _i}+ {\kappa } {\tilde \l _i} {x} )
}{{v_d}} - \frac{( {m^2 _{\tilde L _i }} + 2 {G_+^2} ( {{v_d}}^2 -
{{v_u}}^2 ) + {{\tilde \l _i}}^2 ( {{v_u}}^2 + {{x}}^2 ) )
{v_i}}{{v_d}} \nonumber \\ && - \frac{2 {G_+^2} {{v_i}}^3} {{v_d}},\\
M^2 _{p, uu} &= &\frac{{v_d} {x} ( {A _\l \l } + {\kappa } {\l } {x} )
}{{v_u}} + \frac{{x} ( {A _{\tilde \l _i} \tilde \l _i} + {\kappa }
{\tilde \l_i} {x} ) {v_i}}{{v_u}},\\ M^2 _{p, uS} & = & {v_d} ( {A _\l
\l } - 2{\kappa } {\l } {x} ) + ( {A _{\tilde \l _i} \tilde \l _i} - 2
{\kappa} {\tilde \l _i} {x} ) {v_i}, \\ M^2 _{p, u\tilde \nu _i } & =
& {x} ( {A_{\tilde \l _i} \tilde \l _i} + {\kappa } {\tilde \l _i} {x}
), \\ M^2 _{p, S S} & = & (4 {\kappa } {\l } {v_d} {v_u} + \frac{{A
_\l \l }{v_d} {v_u}} {{x}} + 3 {A _\kappa \kappa } {x}) + \frac{{v_u}
( {A_{\tilde \l _i} \tilde \l _i} + 4 {\kappa } {\tilde \l _i} {x}
){v_i}}{{x}},\\ M^2 _{p, S\tilde \nu _i } & = & {v_u} ( {A _{\tilde
\l_i} \tilde \l _i } - 2 {\kappa } {\tilde \l _i} {x} ),\\
M^2_{p,\tilde \nu _i \tilde \nu _i } & = & {m^2 _{\tilde L _i }} + 2
{G_+^2}( {{v_d}}^2 - {{v_u}}^2 ) + {{\tilde \l _i }}^2 ( {{v_u }}^2 +
{{x}}^2) + 2 {G_+^2 } {{v_i}}^2 .  \eea The above formulae agree in
the case of vanishing sneutrino VEVs, $ v_i =0$, with those obtained
in our previous work~\cite{Chemtob:2006ur}.

The change of basis to the would-be Goldstone bosons in charged and
neutral sectors is implemented through the transformation \bea
{G^+\choose h^+} & = & \calr _\b { H_d^{-\star } \choose H_u^+},\, \,
{G^0 \choose A ^0} = \calr _\b { H_{dI} \choose H_{uI} },
\label{gboson}\\ 
\calr _\b & = & \pmatrix{\cos \b & -\sin \b \cr \sin \b & \cos \b }.
\eea In order to project out the CP-odd modes and the massless
Goldstone mode, $G ^0$, one simply needs to apply on the 2-dimensional
Higgs bosons subspace the $2\times 2 $ matrix rotation
$(H_{dI},H_{uI}) ^T \to (G^0 ,\ A ) ^T = \calr _\b (H_{dI}, H_{uI}) ^T
$, where the label $I$ refers to the imaginary part of the fields,
namely $ H_{d I} = \sqrt {2} \Im (H_{d} )$.  The gauge basis for the
physical CP-odd scalar fields is given by $ (A, \Im (S ) , \Im (\tilde
\nu _i )) ^T = \hat \calr _\b (H_{dI}, H_{uI}, S_I ) ^T$, where $ \hat
\calr _\b = {\mathrm diag} (\calr _\b , 1)$.  In the basis with $ v_i
\ne 0$, the zero eigenvalue Nambu-Goldstone neutral and electrically
charged bosons, defined by the conditions $ M^2 _{p, ij}G^0 _j =0$ and
$ M^2 _{ch, ij} G^+ _j =0$, are given in our current choice of basis
by the column vectors $ G^0 = (v_d , -v_u, 0 , v_i ) ^T$ and $ G^- =
(v_d , -v_u, v_i ,0) ^T$.  The mass eigenstate fields are defined in
terms of the mixing matrices which diagonalize the squared mass
matrices as \bea && (\Phi _ {s, p; I})_{mass} = U_{s,p; Ii} \Phi _
{s,p; i},\ [U_{s,p} M^2 _{s,p} U_{s,p} ^T= (M^2 _{s,p})_{diag} ].
\eea Among the variety of observables of use in testing the scalar
sector (partial decay rates $ Z\to H_I + H_J,\ Z\to H_I + A_J,\ Z\to
\tchi _l ^0 + \tchi _m ^0 $) we mention, for later reference, the
ratio of $Z$ boson vertex couplings $ZZH_I$ to the Higgs bosons
defined by $\xi _{ZZH_I} = ({g _{ZZH_I} \over g _{ZZh} ^{SM} } )^2 =
[(U_s ^{T})_{1I} \cos \b + (U_s ^{T})_{2I} \sin \b ] ^2 $.

\subsection{Mass Matrices  for Neutralinos and Charginos} 

The mass matrix for the coupled system of neutralino and neutrino
fields has been calculated in our previous
paper~\cite{Chemtob:2006ur}.  Since the formulae for this matrix are
rather complicated and were given in full form there, we do not
reproduce these results here.  We shall now concentrate on the mass
matrix for the charginos/charged leptons.  The mass term in the
\underline{Lagrangian} for the charginos receives contributions coming
from the following sources:

1. Contributions from gauge interactions:

\bea ig\sqrt 2 T^a_{ij} [ \lambda^a \psi_j\phi_i^*+ H.c], \eea where
$T^a$ is the generator of the underlying gauge group, $\lambda^a$ is
the corresponding gaugino, and $\phi_i, \psi_j$ are the components of
the matter superfield.

2. Contributions from the superpotential~(in our sign convention):

\bea +\frac{1}{2} \left[\frac{\partial^2 W}{\partial \phi_i \partial
\phi_j} + H.c.\right], \eea where W is the superpotential, and
$\phi_i$ are the scalar components of a chiral superfield.

3. Soft supersymmetry breaking gaugino masses: \bea \frac{1}{2}M_2
\sum_i \tilde\lambda_2^i \tilde\lambda_2^i, \eea where``2'' here
refers to the $SU(2)_L$ gauge group, and $i$ are $SU(2)_L$
indices. Putting together all these contributions, we can write the
mass term for the charginos/charged leptons as \bea L _{mass} & = &
-\frac{1}{2} (\tchi ^{+T}, \tchi ^{-T}) \pmatrix{0 & M_{\tchi}^T \cr
  M_{\tchi} & 0} \pmatrix{\tchi^{+}\cr \tchi^{-}}, \eea where we have
chosen the basis \bea \tchi^{+T} & = & ( -i \tilde \lambda ^+, \tilde
H_u^+, e_R ^{+ }, \mu_R^{+}, \tau_R^{+}), \\ \tchi^{-T } & = & ( -i
\tilde \lambda^-, \tilde H_d ^-, e_L^{-}, \mu_L ^{-}, \tau_L^{-}),\\
\tilde \lambda^{\pm} & = &\frac{1}{\sqrt 2} (\tilde \lambda_2^1 \mp i
\tilde \lambda_2^2), \eea with the mass matrix for the
charginos/charged leptons given by \bea M^2 _{\tchi } & = & \pmatrix{
  M_2 & g_2 v_u & 0 & 0 & 0 \cr g_2 v_d & \lambda x &
  (\lambda^e)_{11}v_1 & (\lambda^e)_{22}v_2 & (\lambda^e)_{33}v_3 \cr
  g_2 v_1 & \tilde \lambda_1 x & -(\lambda^e)_{11} v_d & 0 & 0\cr g_2
  v_2 & \tilde \lambda_2 x & 0 & -(\lambda^e)_{22} v_d & 0\cr g_2 v_3
  & \tilde \lambda_3 x & 0 & 0 & -(\lambda^e)_{33} v_d \cr} . \eea
\label{charginomass}
Here we have assumed that the lepton Yukawa coupling and mass
matrices, $\lambda ^e_{jk} = - M^e _{jk} / v_d $, are diagonal.

\section{Implications  on  Scalar Sector  and   Vacuum Stability}
\label{sec3} 

Before discussing the scalar sector, we briefly discuss some numerical
predictions for the light neutrino masses.  The tree level
contribution and the supposedly dominant one-loop contribution
arising from neutralino-slepton exchange, give a neutrino mass matrix
of approximate structure in the lepton generation space, $ m_{\nu , ij
} \simeq a \tilde \l '_i \tilde \l '_j + b \eta _i \eta _j , \ [\tilde
\l _i ^{'} = \tilde \l _i -\l v_i/v_d ,\ \eta _i = B'_i /B = B_i/B -
v_i /v_d ]$ where the coefficients $a,\ b $ are expected to have a
smooth dependence on the parameters $\l,  \ \tan \b$, and $ x $.  The
pair of non-vanishing mass eigenvalues of this matrix are expressed in
terms of the invariant parameters by \bea && m_ {\nu _2} = {b \over 2
\tilde \l ^{'2} } \vert \vec {\tilde \l '} \wedge \vec \eta \vert ^
2,\ m_ {\nu _3} = a \tilde \l ^{'2} + {b\over \tilde \l ^{'2} } (\vec
{\tilde \l ' }\cdot \vec \eta ) ^ 2, \cr && [ \tilde \l ^{'2} = \sum
_i \tilde \l ^{'2} _i,\ \eta ^{2} = \sum _i \eta ^{2} _i, \ \vert \vec
{\tilde \l '} \wedge \vec \eta \vert ^2 = \ud \sum _{i,j} (\tilde \l '
_i \eta _j - \tilde \l ' _j \eta _i)^2,\ \vec { \tilde \l '} \cdot
\vec \eta = \sum _i \tilde \l ' _i \eta _i ] .\eea For clarity, we
quote the explicit expression for the single non-vanishing neutrino
Majorana mass eigenvalue present at the tree
level~\cite{Chemtob:2006ur} \bea && m_ {\nu _3} = {\s x^2 V_u ^2 (
M_2/M_1+ g_2 ^2/g_1 ^2 ) (\tilde \l _i - v_i \l / v_d )^{2} \over \tan
^2 \b \bigg [ V_u ^2 ( M_2/M_1 + g_2 ^2/g_1 ^2 ) \bigg ( 4 \kappa \l x ^2
/\tan \b + (\rho /\tan \b - \l v_u )^2 \bigg ) - M_2 \l ^2 x^2 \s
\bigg ] }
, \cr && [\s = 2( \kappa x - {\rho v_u \over x }),\ \rho = -\sum _A
\tilde \l_A v_A ,\ \mu =- \l x ,\ V_u ={g_1 v_u \over \sqrt 2} ]
. \label{eqneutmass} \eea

We have attempted to extract the values of the coefficients $ a$ and
$b$ from numerical calculations of the tree level formula,
Eq.~(\ref{eqneutmass}), and of the one-loop amplitude in Eq.~(III.54)
of our previous work~\cite{Chemtob:2006ur}. Using the reference set of
parameters given in the caption of Fig.~\ref{specscalar} with $\tan
\b =2$, gives us the following order of magnitude estimates: $ a
\approx 10 \ \text{GeV},\ b \approx 10 ^{-1} $ GeV.  For these
predictions to be compatible with the limit on the neutrino mass, $
\vert m_{\nu }\vert < 1 $ eV, we need to impose $ \l _i ^{'} < 10^{-5}
,\ \eta _i < 10^{-4} .$ We have not attempted a global fit to the
neutrino experimental data but note that this could be done by
following a similar treatment as that used for the MSSM with bilinear
R parity violation~\cite{chun02}.

\subsection{Numerical Study of  Scalar Sector}

\begin{center} \begin{figure}[htb] \epsfxsize =5.in
\epsfysize=5.in \epsffile{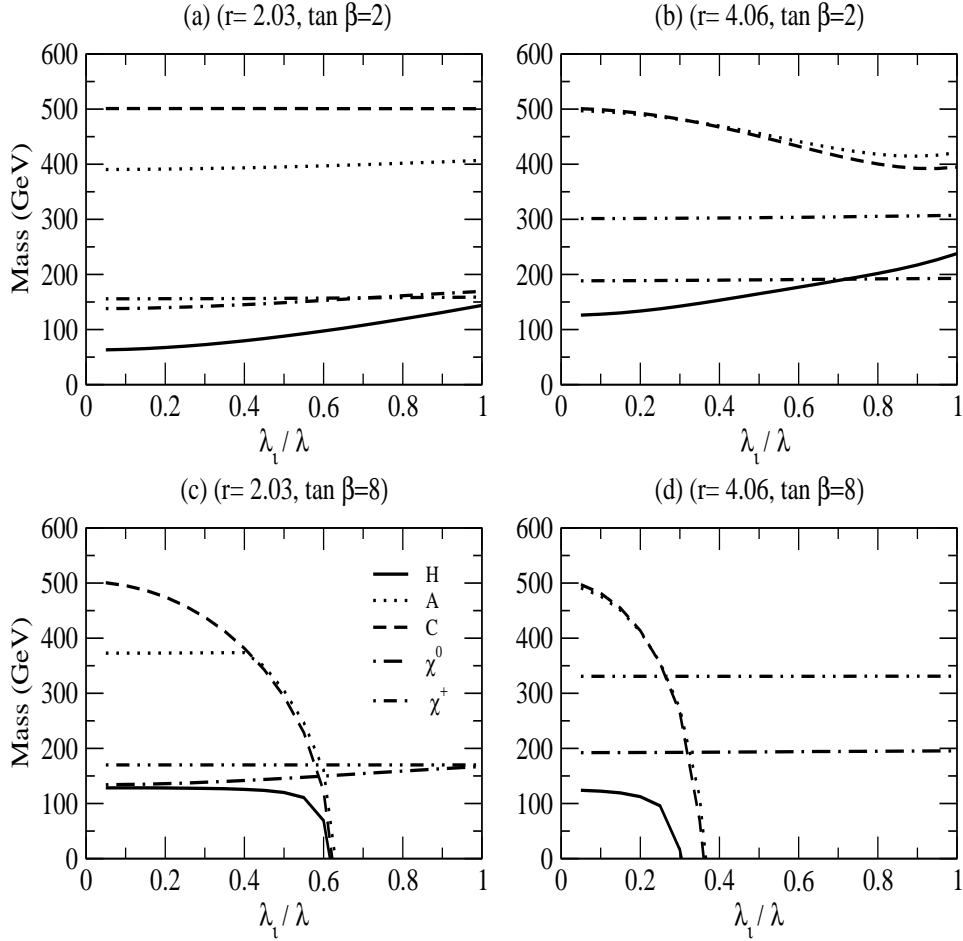}
\caption{The masses of lowest lying CP-even and CP-odd neutral
scalars, the charged scalars, and the neutralino and chargino states
are plotted as a function of $\tilde \l _i / \l $.  We assume the
single lepton number violating dominance hypothesis with a fixed
generation $\tilde \l _i \ne 0 $ considered one at a time.  The curves
for $H_{I=1} ,\ A_{I=1},\ C_{I=1},\ \tchi ^0_ {l=1} $ and $\tchi
^+_{l=1} $ are drawn in full, dotted, dashed, dot-dash and double-dot
dashed lines, respectively, as illustrated in the legend.  The upper
panels $(a) $ and $ (b) $ correspond to $\tan \b = 2$ whereas the
lower panels $(c) $ and $ (d) $ to $\tan \b = 8$. The left hand panels
$(a)$ and $(c) $, and the right hand panels $(b)$ and $(d) $ refer to
the parameter $ r \equiv x/v = 2.03 $ and $ 4.03 $, respectively. Our
choice of the parameters, as given in Ref.~\cite{Barger:2006dh}
(Fig. $1$), is described by the values: $\l =0.5 ,\ \kappa= +0.5 ,\
A_\l =500 \ \text{GeV},\ A_\kappa=+250 \ \text{GeV} $, and $ A^u_t= 1
\ TeV , \ m _{\tilde Q } = m _{\tilde U^c } = 1 \ TeV, \ Q = 300 \
\text{GeV} $, which enter the tree and one-loop contributions to the
scalar potential.  In the prescription I in which we work, we set the
sneutrino mass parameter at $m^2_{\tilde L_i} = 500 \ \text{GeV}$.
The neutralino-neutrino and chargino-lepton mass matrices are
evaluated by assuming the relation between the gaugino mass parameters
$ M_1 = { k_1 g_1 ^2 \over k_2 g_2^2} M_2 ,$ with $k_1 = {5\over 3} ,
k_2 =1, $ while using the numerical values $M_2 = 400 \ \text{GeV}$
and $ M_1 = 199.2 \ \text{GeV}.$ The values of parameters $v_i$ and $A
_{\tilde \l _i} $ are set through the alignment conditions $\tilde \l
' _i \approx 0 ,\ \eta _ i \approx 0 $, where $\tilde \l ' _i \equiv
\tilde \l _i - {\l v_i \over v_d } ,\ \eta _ i \equiv { A _{\tilde \l
_i} \tilde \l _i \over A_\l \l } - {v_i \over v_d} \equiv { B'_i \over
B} ,\ [B'_i=B_i - B {v_i \over v_d},\ B_i =A _{\tilde \l _i} \tilde \l
_i ,\ B= A_\l \l ] $.}
\label{specscalar} \end{figure} \end{center}

\begin{center} \begin{figure}[htb] \epsfxsize =5.in
\epsfysize =5. in \epsffile{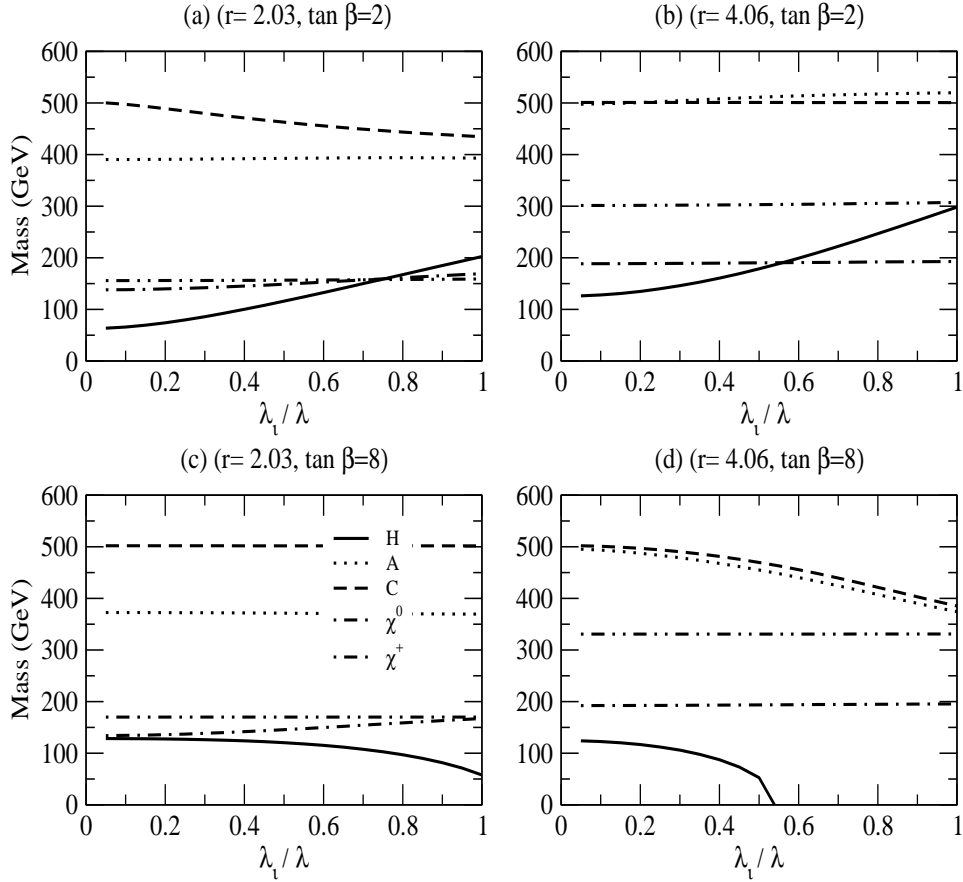}
\caption{The masses of the lightest CP-even and CP-odd neutral scalars
are plotted as a function of $\tilde \l _i / \l $, for fixed lepton
generation, under same conditions as in Fig.~\ref{specscalar}.  The
single lepton number violating dominance hypothesis is assumed such
that finite coupling constants $\tilde \l _i$ of fixed generation are
considered one at a time.  We use the same set of NMSSM parameters
as in  Fig.~\ref{specscalar}, the only change being the vanishing 
trilinear supersymmetry breaking coupling, $A _{\tilde \l _i} =0$.  
We have also displayed the lightest
neutralino and chargino masses although these are independent of the
\susy breaking parameter $A _{\tilde \l _i},$ and are given by the
same curves as in Fig.~\ref{specscalar}.}
\label{spec2}  \end{figure} \end{center}

\begin{center} \begin{figure}[htb] 
\epsfxsize =5.in \epsfysize =5.in \epsffile{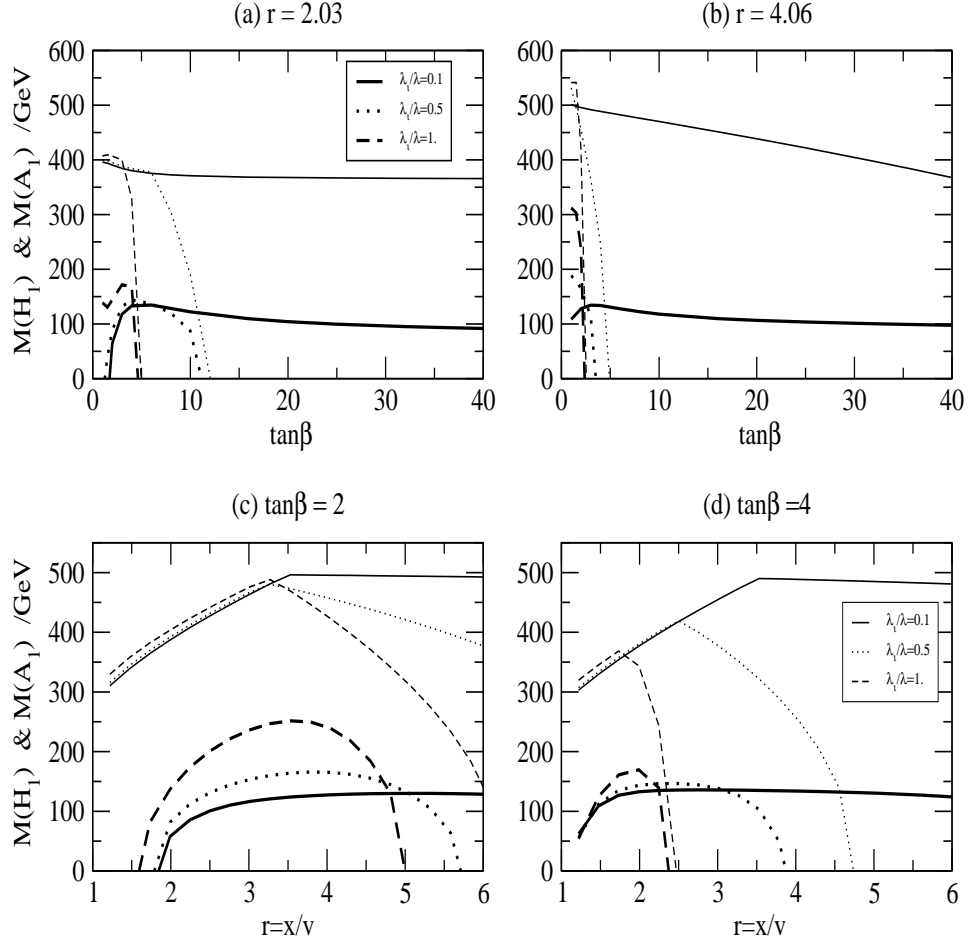}
\caption{The masses of the lightest CP-even and CP-odd neutral scalars
plotted as a function of $\tan \b $ at two fixed values of $r = x/v =
2.03 $ and $ 4.06 $ (upper panels $(a)$ and $ (b) $), corresponding to
$ x\sqrt 2 = 500 \text{GeV} $ and $ 1 $ TeV, respectively, and as a
function of $ r = x/v$ at the two fixed values of $ \tan \b = 2 $ and
$ 4 $ (lower panels $(c) $ and $ (d) $).  We use the same set of
parameters as in Fig.~\ref{specscalar} and assume the single lepton
number violating dominance hypothesis with a finite $\tilde \l _i$ of
fixed generation.  The three curves for the modes $H_{I=1} $ (bold)
and $ A_{I=1} $ (light) are drawn in full, dotted and dashed lines
corresponding to the values $\tilde \l _i /\l = 0.1,\ 0.5, \ 1$.}
\label{figtanb}  \end{figure} \end{center}

In order to have a reliable semi-quantitative description of the
NMSSM, we have added to the tree level scalar potential the lepton
number conserving one-loop contributions from the top quark and stop
pair of squarks. The corrections to the neutral scalar  mass
matrix are obtained by evaluating Eq.~(\ref{eqcolwei}) using a
standard procedure~\cite{Barger:2006dh}, so we refrain from quoting
the explicit formulas.  The results of our numerical calculations for
the mass spectra of the scalar and sfermion modes are shown in
Figs.~\ref{specscalar}, \ref{spec2} and \ref{figtanb} as single
parameter plots displaying the dependence on the lepton number
violating couplings $\tilde \l _i $.  These results were obtained by
considering the reference set of natural input parameters used in 
~\cite{Barger:2006dh},  which are detailed in the
caption of Fig.~\ref{specscalar}.  Although our analytical formulae
include the trilinear lepton number violating interactions with
coupling constants $ \l _{ijk},\ \l '_{ijk}$, in our numerical
calculations we have retained only the lepton number violating
couplings $\tilde \lambda_i$.

In Fig.~\ref{specscalar} we present the results of our numerical
calculations for the masses of the lightest neutral and charged scalar
modes, and neutralino and chargino modes as a function of the lepton
number violating coupling, which we take to vary within the range 
$\tilde \l _i / \l \in [0, \ 1]$.  At small $\tan \b = O(1) $, 
the modifications induced
by $\tilde \l _i$ are essentially quantitative.  As the coupling
increases, the lightest CP-even scalar mass is enhanced and the CP-odd
and charged scalar masses are reduced. This effect of enhancement and
reduction of masses are clearly seen at large values of $ r \equiv
x/v$.  These results are also sensitive to the value of the VEV ratio
$ \tan \b = v_u/v_d $.  While the constraints on the scalar masses
restrict $ \tan \b $ to a small natural range in the NMSSM, this
feature is even more pronounced for large values of the lepton number
violating interactions $\tilde \l _i \approx \l .$ Already for the
value $\tan \b = 8$, the mass spectra undergo a qualitative change
with large negative contributions to both the neutral and charged
scalar bosons causing the vacuum instability signalled by the
occurrence of tachyons at $\tilde \l _i / \l \geq 0.5 $.  This is
explained by inspection of Eq.~(\ref{eqmatchscale}) which shows how
increasing $ v_i / v_d$ proportionately to $\tilde \l _i / \l $ has
the same effect as an increase of $\tan \b $.  The allowed intervals
for $ \tan \b $ further shrink with increasing $x$.  We also present
in Fig.~\ref{specscalar} the masses of the lowest lying neutralino
and chargino states.  The results are characterized by a slow growth
of the masses with increasing $ \tilde \l _i / \l $.

The coupling of $Z$ to the Higgs, $ \xi _{ZZH_1},$ is found to vary in
the intervals $ (0.84 \to 0.63) , \ (0.99 \to 0.91),\ (0.96 \to
0.81),\ (0.99 \to 0.94)$ as $\tilde \l _i /\l $ increases in the the
range $ (0.1 \to 1.0)$ for the mass spectra displayed in panels $(a),\
(b),\ (c),\ (d)$ of Fig.~\ref{specscalar}.  For the sake of a
qualitative comparison with experiment, we note that the LEP data
indicates a correlation between the lightest Higgs boson vector
coupling and mass such that $ \xi _{ZZH_1} < (0.02 \ - \ 1.0 ) $ for $
M_{H_1} = (50 \ - \ 120) $ GeV.

In order to understand the dependence of the scalar mass spectrum on
the \susy breaking interactions, we display in Fig.~\ref{spec2} the
results obtained for the case $ A _{\tilde \l _i } =0$.  This choice
is made only for illustrative purposes since it violates the alignment
condition for the parameters $\eta _i$.  As compared to our initial
choice, $ A _{\tilde \l _i } \ne 0$, in Fig.~\ref{specscalar}, we see
that switching off the \susy breaking trilinear lepton number
violating interactions results in an enhanced effect at small $\tan \b
\simeq 2$ but a reduced effect at $\tan \b \simeq 8$. Indeed, the
vacuum instability at $\tan \b =8 $ sets in at larger values of $
\tilde \l _i /\l $.

In Fig.~\ref{figtanb} we display one-parameter plots for the masses of
the lowest lying neutral CP-even and CP-odd Higgs bosons as a function
of $\tan \b $ (fixed $x$) and as a function of $x$ (fixed $\tan \b $)
for the discrete set of values for $\tilde \l _i / \l = 0.1, \ 0.5,\
1. $  From these results we see that the maximal values of $\tan \b $
allowed by vacuum stability lie inside the range $ 5\ -\ 10 $, and
that this interval gets further reduced at larger values of the VEV of
the singlet field $x$. As a related observation, we note that the
allowed interval for $r \equiv x/v \in [1,6] $ shrinks to $r \in
[1,2]$ at $\tan \b =4$ for large enough values of $\tilde \l _i $.
Furthermore, we observe that for the fixed values of $ r = 2.03 ,\
4.06$ in the panels $(a) , \ (b)$, corresponding to $ x \sqrt 2 = 500
\ $ GeV and $1 $ TeV, respectively, our results at small $\tilde \l _i
/ \l \simeq 0.1 $ join smoothly with those at $\tilde\l _i =0 $, as
can be seen in Fig. $1$ (panels $(a) $ and $(c)$) of
Ref.~\cite{Barger:2006dh}.

\vskip 1 cm 
\vskip 1 cm

\vskip 1 cm 
\vskip 1 cm

\vskip 1 cm

\section{Constraints   from Unbounded from Below Directions
and Charge and Color Breaking Minima}
\label{sec4}

\subsection{General  Considerations}

The stability of the regular vacuum with respect to the UFB directions
and the distant CCB minima in the scalar field space are expected to
give useful constraints on the NMSSM parameters.  We briefly recall
here the main features of the renormalization group approach to this
problem which was extensively discussed in the context of
MSSM~\cite{Casas:1997ze}. This approach is particularly useful in the
context of a gravity mediation of \susy breaking in grand unified
theories~(supergravity GUT), where the unknown independent parameters
consist of a small set of parameters and one can develop semi-analytic
methods.  The idea is to select suitable directions in the scalar
field space along which the contributions to the scalar potential are
dominated by the \susy breaking terms while those from $F-$ and $D-$
terms are smaller and smoothly varying.  In practice, the field
directions are parameterized in terms of some single real variable,
$w$, along which the scalar potential is given by a polynomial in $w$
which receives, for a suitable choice of the parameter phases,
negative contributions of $O(w^2)$ or $O(w^3)$ from the \susy breaking
trilinear scalar couplings and mass terms, and positive contributions
of $O(w^4)$ or higher order from the \susy terms.  The higher order
positive terms lift the potential upwards at large $w$.  Even when
absent at the tree level, higher order contributions to the potential,
which lift it, always arise at the one-loop order.  Requiring the
scalar potential along these directions to stay above the regular
minimum yields algebraic conditions on the coefficients of the
polynomial. Considering, for illustration, the case of the field
direction yielding the quartic order potential, $ V (w ) = A w ^4 + B
w ^2 + C,$ where the coefficients $ A, \ B, \ C$ are known algebraic
functions of the various parameters, we see that this is an UFB
direction if $ A=0$. At $A>0$, this direction develops a minimum if $
B<0$, as seen by evaluating the extremum with respect to $w^2$, \bea
&& w _{ min} ^2 = - { B \over 2 A } \ \ \Longrightarrow \ \ V (w _{
min} ) = - { B^2 \over 4 A } + C . \eea The regular electroweak
symmetry breaking vacuum is stable against decay to the new vacuum
solution as long as $V (w _{ min} ) > V_{MIN}$, where $ V_{MIN}$
denotes the value at the regular minimum.  An improved quantum
stability condition can be deduced without a detailed knowledge of the
one-loop radiative corrections to the potential upon invoking the
logarithmic dependence on the masses of states scaled by the running 
momentum scale $Q$.  One needs only to assume that the above
inequality involving the tree level scalar potential still holds but
with the parameters replaced by renormalization group improved,
momentum scale $Q $ dependent, running parameters, \bea V _{UFB} (w ;
Q = \hat Q ) & > & V_{MIN} ( w _0 ; Q = M_S ),
\eea where the arguments $w$ and $Q$ along the field direction are set
in the grand unification framework as $ w \in [ m_W, M_X]$ and $\hat Q
= {\mathrm Max}~ (g_2 w , \l _t w , M_S),$ and those for the regular
minimum are set as $ w_0 = [v_u,\ v_d ]$ and $ Q= M_S $, where $ v_u $
and $v_d $ denote the the usual Higgs VEVs, and $ M_S$ the
effective \susy breaking mass scale.  Extensive analyses of the vacuum
stability constraints have been developed for the
MSSM~\cite{Casas:1997ze}.  For the MSSM with broken R-parity, Abel and
Savoy~\cite{Abel:1998ie} inferred lower bounds on the trilinear
R-parity violating couplings $\l _{ijk} ,\ \l '_{ijk}$, and Hirsch et
al.,~\cite{Hirsch:2004hr} discussed the implications on the bilinear
R-parity violating interactions.

Although a similar programme for the NMSSM with lepton number
violation appears to be well motivated, its implementation is
substantially complicated by the need to consider two-dimensional
field directions which include the singlet field $S=x$ as an
independent variable, since there are no obvious correlations between
the contributions from the electroweak singlet and non-singlet fields.
Another complication stems from the fact that the renormalization
group formalism is more complicated.  This can be seen from the
absence of studies of the dangerous field directions for the NMSSM
beyond the preliminary works~\cite{derensavoy84,ibanmas87,gunion88}.
The progress achieved through the quasi fixed point solution for the
running parameters~\cite{nevzorov1,nevzorov02} or the numerical
studies of physical constraints~\cite{nmssm95,nmssm96,ellwang04} are
not of direct help to us in the present work.  Since our main focus is
on the lepton number violation, rather than solving the full-fledged
problem, we shall follow a simple phenomenological approach, which we
now describe.  We assign natural values for the lepton number
conserving Yukawa couplings and soft parameters which respect the
regular minimum stability, and then examine the effect of increasing
the lepton number violating parameters $\tilde \l _i , \ B_i , \ v_i
$, consistently with the constraints from the neutrino mass matrix.
This approach is similar in spirit to that followed
in~\cite{Hirsch:2004hr}.

The field directions involve suitable subsets of the electrically
neutral and charged scalar states and the squark states.  We continue
using the single lepton flavor  dominance hypothesis in which
the nonvanishing parameters, $\tilde \l _i,\ A_{\tilde \l _i} , \ v_i
$ are finite only for fixed generation labels $i ,\ j , \cdots $.
Useful intermediate formulas for the scalar potential of the
electrically neutral fields and for the classical equations of motion
of the electrically neutral and charged fields are given in
Eq.~(\ref{potneut1}), Eq.~(\ref{eomneut}) and Eq.~(\ref{eomch}) of the
Appendix~\ref{appexa}.

To start with, we need the potential at the regular minimum.  With our
prescription of eliminating the dependence on the soft mass parameters
$ m^2_{H_u} ,\ m^2_{H_d},\ m^2_{ S},\ $ and $ m^2_{ H_d \tilde L_i} $
by using the minimization equations with respect to $ v_u,\ v_d , \ x
,$ and $ v_i $, the value of the scalar potential at the minimum is
given by \bea V_{MIN}(v_u, v_d, x, v_i) & = & - { G_+^2 } {( {
{v_d}}^2 - { {v_u}}^2 ) }^2 - {1\over 3} \bigg [ -6 {\kappa } {\l }
{v_d} {v_u} { {x}}^2 + 3 { {\l }}^2 ( { {v_u}}^2 { {x}}^2 + { {v_d}}^2
( { {v_u}}^2 + { {x}}^2 ) ) \nonumber \\ & + & {x} ( -3 {A _\l \l }
{v_d} {v_u} + {x} ( 3 { {\tilde \l _i }}^2 { {v_u}}^2 - {A_\kappa }
{x} + 3 { {\kappa }}^2 { {x}}^2 ) ) \bigg ] \nonumber \\ & + & \bigg [
{v_u} {x} ( {A_{\tilde \l _i} \tilde \l_i } + 2 {\kappa } {\tilde \l
_i } {x} ) - 2 {\l } {\tilde \l _i } {v_d} ( { {v_u}}^2 + { {x}}^2 )
\bigg ] {v_i} \nonumber \\ & + & \bigg [ 2 G_+^2 ( -{{v_d}}^2 + {
{v_u}}^2 ) - { {\tilde \l _i }}^2 ( { {v_u }}^2 + { {x}}^2 ) \bigg ] {
{v_i}}^2 - G_+^2 {{v_i}} ^4.  \eea The above formula is seen to be
independent of the slepton-Higgs boson soft mass parameters, so that
it holds in the same form in the prescription where one eliminates $
m^2_{ \tilde L_i} $ rather than $m^2_{ H_d \tilde L_i} $.  We also
note that the quartic and quadratic dependence on the VEVs coming from
the gauge interactions combine into a term of same form as the minimum
value of the scalar potential in the MSSM with bilinear $R$-parity
violation \bea (V_{MIN} )_{MSSM} & = & -G_+^2 {( { v_i ^2 + {v_d}}^2 -
{ {v_u}}^2 ) }^2.  \eea
\subsection{Unbounded From Below Directions}

The Higgs-slepton field directions of interest are those which
minimize the positive contributions from $F-$ and $D-$terms and
maximize the negative contributions from soft masses and trilinear
couplings.  There are three main unbounded from below directions
defined by \bea && UFB-1: \ H_d = H_u \ne 0 , \ S \ne 0 ; \cr && UFB
-2: \ H_d \ne 0 ,\ H_u \ne 0 ,\ \tilde \nu _i\ne 0 , \ S \ne 0 ; \cr
&& UFB -3: \ H_d = 0, H_u \ne 0 ,\ \tilde \nu _i \ne 0 , \tilde e_j =
\tilde e^c_j \ne 0 , \ S \ne 0 , \ [i\ne j ].  \eea The singlet field
dependence of the potential makes an analytic study intractable.  For
instance, minimizing the potential with respect to $S$ introduces a
non-polynomial dependence with respect to the variables describing the
electroweak non-singlet field directions.

We now discuss the three UFB directions in detail. Along the $ UFB-1 $
direction described by the two variables $ \vert H_d \vert=\vert H_u
\vert = w $ and $S =x$, the potential is given by \bea V_{UFB-1} (w,
x) & = & A w ^4 + B (x) w ^2 + C (x), \nonumber \\ A & = & \l ^2,\
B(x)= m^2 _{H_u } + m^2 _{H_d} - 2 A_\l \l x + 2 \mu ^2 _{du} + x^2 (
2 \l (\l -\kappa ) + \tilde \l _i ^2 ), \ \nonumber \\ C (x) & = &
\kappa ^2 x ^4 - {2\over 3} A_\kappa \kappa x^3 + m^2 _{S} x ^2 .
\eea We have included the dependence on $\mu ^2 _{du}$ although this
parameter is expected to be absent, as already discussed, in the
minimal version of the model.  For $\l \ne 0$, the would-be UFB
direction is lifted at large $w$, and features a minimum at $ w _{min}
$ unless $ B(x) >0$.  It is useful to note that $ B(x)$ is a quadratic
form in $x$ with an extremum at $x_{min}$ defined by $ \dh B / \dh x
\vert _{x= x_{min} } =0 \ \Longrightarrow \ x_{min} = A_\l \l / [2
\l(\l -\kappa ) + \tilde \l _i ^2 ] $.  Since this is a minimum
provided that $ 2 \l (\l -\kappa ) + \tilde \l _i ^2 >0$, one can
express the condition for the absence of the UFB-1 direction by the
appoximate bound on the soft parameters \bea B(x) & \geq & B(x _{min}
) = m^2 _{H_u } + m^2 _{H_d} - { ( A_\l \l)^2 \over 2 \l (\l -\kappa
)+ \tilde \l _i ^2 } + 2 \mu ^2 _{du} > 0.  \eea It is interesting to
compare with the corresponding bound in the MSSM, $B (x_{min}) \to m^2
_{H_u } + m^2 _{H_d} + 2(\mu ^2 - (A _\mu \mu ) ^2  ).$ To obtain the
improved constraint, one should determine the position of the
potential minimum, $x _{min}$ and $\ w _{min}$, and require the
condition $V _{UFB-1}(x _{min}, w _{min}; \hat Q) \geq V_{MIN}(x =x_0,
v_u = v \sin \b , v_d = v \cos \b ; Q=M_S ).$

The UFB-2 field direction can be conveniently described by the
parameterization $ v_u = w,\ v_d = w \cos \t , \ v_i = w \sin \t , \
x ,\ [\t \in [0, 2 \pi ] ] $, which is designed to cancel the
$D-$terms.  The scalar potential is given by the quartic order
polynomial in $w$ \bea V _{UFB-2} (w, x , \t ) & \equiv & A (x, \t )
w^4 + B(x, \t )w^2 + C(x, \t ), \eea where \bea A (x, \t ) & = & { (
{\l }\,\cos ( {\t }) + {\tilde \l _i}\,\sin ( {\t }) ) }^2, \\ B(x, \t
) & = & { {\l }}^2\,{ {x}}^2 + { {\tilde \l _i}}^2\,{ {x}}^2 + {m^2
_{H_u}} - 2\, {x}\, ( {A _\l \l } + {\kappa }\, {\l }\, {x} ) \, \cos
( {\t }) \nonumber \\ & + & ( { {\l }}^2\,{ {x}}^2 + {m^2 _{H_d} } )
\,{\cos ^2( {\t })} - 2\, {x}\, ( {A _{\tilde \l _i} \tilde \l _i } +
{\kappa }\, {\tilde \l _i}\, {x} ) \, \sin ( {\t }) \nonumber \\ & + &
( { {\tilde \l _i}}^2\,{ {x}}^2 + {m^2 _{\tilde L_i }} ) \, {\sin ^2(
{\t })} + {\l }\, {\tilde\l _i}\,{ {x}}^2\, \sin (2\, {\t }),
\label{eqpufb2} \\ C (x, \t ) & = & \frac{-2\, {A _\kappa \kappa }\,{
{x}}^3}{3} + { {\kappa }}^2\,{ {x}}^4 + { {x}} ^2\, {m^2 _{S}}.  \eea

\begin{center} \begin{figure}[b] 
\epsfxsize =6.in \epsfysize =6.in 
\epsffile[72 172 580 620]{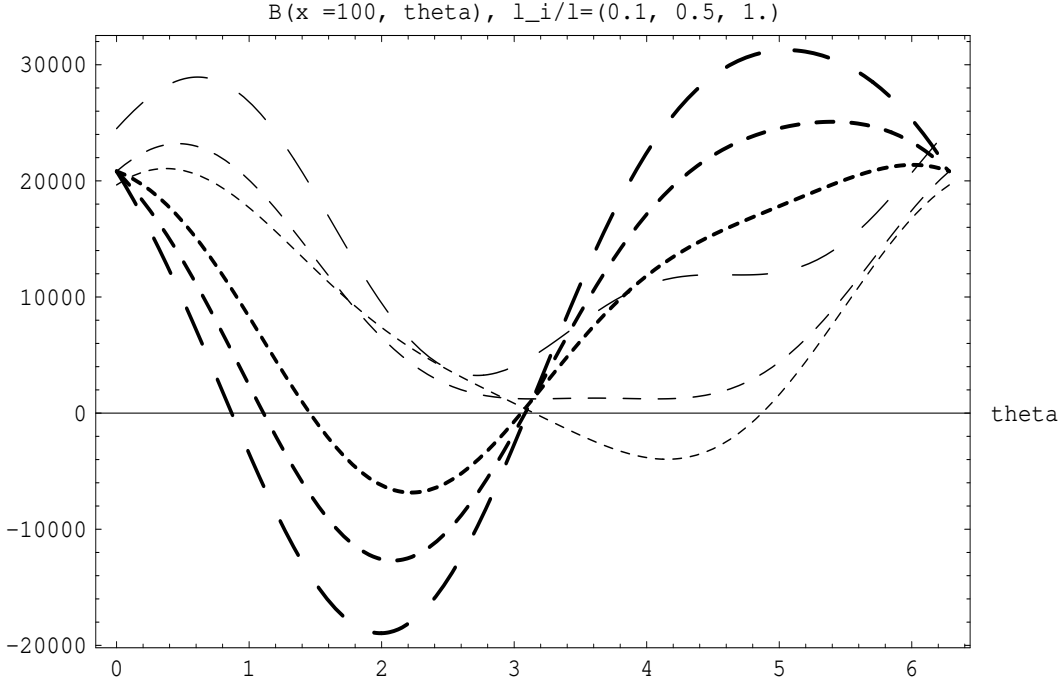}
\caption{The coefficient $ B(x , \t )$ for the UFB-2 direction is
plotted (in $ \text{GeV} ^2$ units) as a function of $ \t $ for the
relevant input parameters set as, $\l =0.7 ,\ \kappa =0.3 ,\ A _\l =
-100  \ \text{GeV} , \ m^2 _{H_d} = m^2 _{\tilde L, i} = + 100 \
\text{GeV}^2 ,\ m^2 _{H_u} =- 100 \ \text{GeV}^2, \ x=100 $ GeV, and a
discrete set of choices for $\tilde \l _i $ and $A_{\tilde \l _i }.$
The three curves for $A_{\tilde \l _i } = - 100 \ \text{GeV} $ in the
three cases $\tilde \l _i /\l = (0.1,\ 0.5,\ 1.) $ are drawn in light
lines with dashes of increasing lengths.  The three curves for $\tilde
\l _i /\l = 0.5 $ in the three cases $A_{\tilde \l _i } = (100 ,\ 200,
\ 300 ) \ \text{GeV} $ are given in thick lines with dashes of
increasing lengths.}
\label{figufb2}  \end{figure} \end{center}

Since the UFB-2 direction coincides with the UFB-1 direction at $ \t =
0$, to determine whether one avoids  a distant minimum along $w$
it is only necessary to test the condition that the coefficient $ B(x,
\t ) <0 $ at some finite $\t $.  Inspection of Eq.~(\ref{eqpufb2}) for
$ B(x, \t ) $ shows that the condition $ B(x, \t =0 ) > 0 $ is
sensitive to the signs of parameters $ m^2_{ H_d } , \ m^2_{ H_u} $
and $ A_\l $, while the condition  $ B(x, \t ) > 0$ at finite $ \t
$ is sensitive to the signs of $ A _ {\tilde \l _i } $ and $
m^2_{\tilde L_i } $.  To achieve $ B(x, 0) >0$, it is more favorable
to restrict to the choice $ A_\l < 0$.  We now attempt to assign by
hand typical values to the relevant free coupling and mass
parameters, and determine numerically whether the lepton number
violating interactions can drive $B(x, \t ) $ to negative values if it
started from a positive value at $\t =0 $.  The dependence on $\t $ is
displayed in Fig.~\ref{figufb2} for typical values of the input
parameters which are specified in the caption of that  figure. 
We see from these results that for trilinear couplings $ A_{\tilde \l
_i } $ of same negative sign as $ A_\l $, the $\tilde \l _i$
interactions give significant positive sign contributions to $B(x ,
\t)$ which avoid the occurrence of dangerous minima along $w$ at
finite $\t $.  By contrast, choosing triscalar couplings of opposite
sign, $ A_{\tilde \l _i } >0$, always drives $B(x, \t ) $ to negative
values at finite $\t .$ We note that a dependence on $\t $ of similar
type is found with various other choices of $ x$ and that changing the
sign of $ \tilde \l _i $ leaves the potential unchanged up to the
replacement, $ \t \to \pi -\t $.  We, thus, conclude that the lepton
number violating couplings, $\tilde \l _i$ and $ A_{\tilde \l
_i } $, have the ability to remove or induce the dangerous minima  at
finite $\t $    depending on whether $ A_{\tilde \l _i } $
is negative or positive, namely, of same or opposite sign to $A_\l
$.  We  note that the case with   $A_\l $ and $ A_{\tilde \l
_i } $ of opposite signs  clashes with the universality of supersymmetry 
breaking and is unlikeky to occur in the context of
grand unified theories.

\begin{center} \begin{figure}[htb] 
\epsfxsize =6.in \epsfysize =6.in  
\epsffile[72 172 580 620]{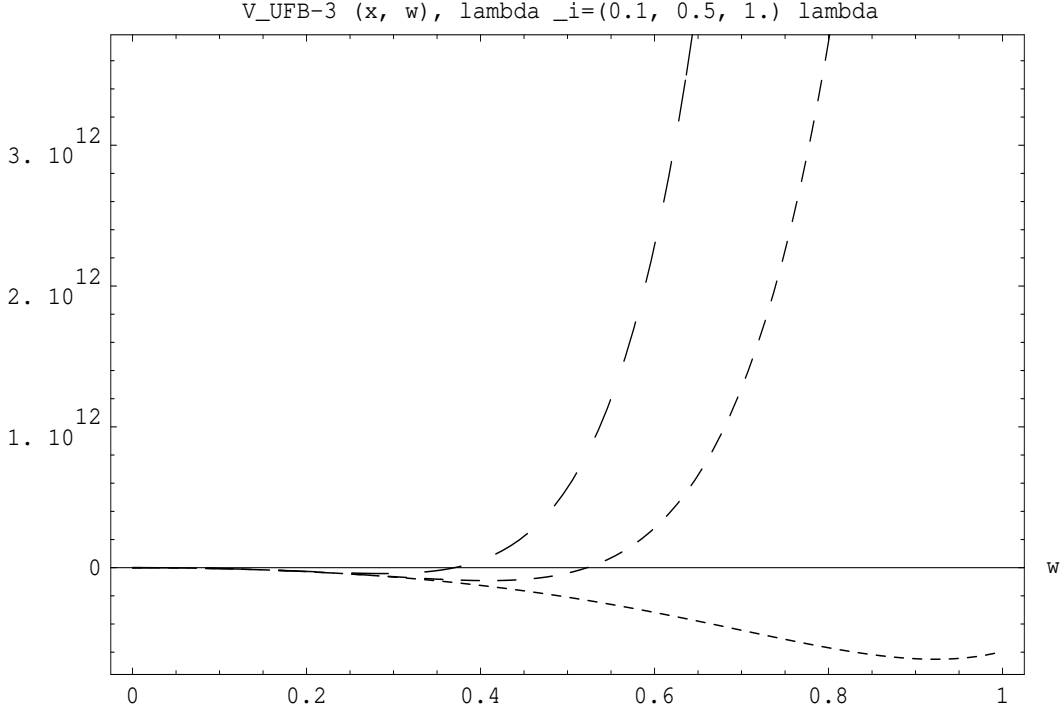}
\caption{The potential energy density $ V(x , w )$ for the UFB-3
direction is plotted in $ \text{GeV} ^4$ units as a function of $ w $
for the choice of input parameters, $\l =0.7 ,\ \kappa =0.3 ,\ A _\l =
500 \ \text{GeV} ,\ A_\kappa =500 \ \text{GeV} , \ A_{\tilde \l _i } =
500 \ \text{GeV} , \ m^2 _{H_u} = m^2 _{H_d} = m^2 _{\tilde L, i} = -
200 \ \text{GeV}^2,\ m^2 _{\tilde E^c _i} = m^2 _{S } = 100 \
\text{GeV} ^2 ,$ with VEV parameters, $ x= 100 \ \text{GeV} ,\ \tan
\b = 2$ and $ \s =\l x / \l ^e _\tau =\l x v_d/ m_\tau ,\ [m_\tau =
1.777 \ \text{GeV} ] $ for the tau-lepton field case.  The reduced
range of variation of the  variable $ w$ is explained by
the large value assumed  by the   scaling  factor, 
$\s \ \approx 30.6 \ x $.  The curves for
$\tilde \l _i /\l = (0.1, \ 0.5,\ 1.) $ are drawn with dashes of
increasing lengths.}
\label{figufb3}  \end{figure} \end{center}

The UFB-3 direction can be conveniently described by the
parameterization $ v_u = \s w ^2, \ e_j = e^c_j = \s w , \ v_i = \s w
(1 + w ^2 ) ^ \ud  \ [i\ne j],$ designed to cancel the $D-$terms,
while using $\s = {\l x \over \l ^e _j} $, in order to satisfy the
$F-$term flatness condition $ W _{H_d} =0$.  The potential along this
direction is given by \bea V_{UFB-3} (w, x) & = & - {2\over 3}
A_\kappa \kappa x^3 + { {\kappa}}^2\,{ {x}}^4 + { {\tilde \l _i}}^2\,{
{w}}^6\,{ {\s }}^4 + { {\tilde \l _i}}^2\,{ {w}}^8\,{ {\s }}^4 + {
{w}}^3\, ( -2\, {A_{\tilde \l _i } \tilde \l _i }\, {\sqrt{1 + {
{w}}^2}}\, {x}\,{ {\s }}^2 \nonumber \\ && - 2\, {\kappa }\, {\tilde
\l _i}\, {\sqrt{1 + { {w}}^2}}\,{ {x}}^2\, { {\s }}^2 ) + { {w}}^2\, {
{\s }}^2\, ( 2\,{ {\tilde \l _i}}^2\,{ {x}}^2\, + {m^2 _{\tilde L_j }}
+ {m^2 _{\tilde E^c_j }} ) \nonumber \\ && + { {w }}^4\, { {\s }}^2\,
( { {\l }}^2\,{ {x}}^2\, + 2\,{ {\tilde \l _i }}^2\,{ {x}}^2\, + {m^2
_{\tilde L_i }} + {m^2 _{H_u}}) + {m^2 _{S} } { {x}}^2 - 2 A^\l _{ijj}
\l _{ijj} \s ^3 w^3 (1+w^2)^\ud \nonumber \\ && + \s ^3 w^4 [ 2 \l
_{Ajj} \tilde \l _A x + \l _{Ajj} ^2 \s + \l _{Aij} ^2 \s w^2 (1+w^2)
+ \l _{ijj} ^2 \s (1+w^2) ].
\label{equfb3} \eea 
Inspection of the above potential indicates that the lepton number
violating contributions from $\tilde \l _i$ or $ \l _{ijk}$ add
positive terms to the potential that have the ability to lift the
UFB-3 direction.  We restrict ourselves to an illustrative example by
assigning the NMSSM parameters typical values resulting in a potential
with a deep minimum at $\tilde \l _i =0$, and determining whether
switching on $\tilde \l _i$ to finite values lifts this minimum.  In
Fig.~\ref{figufb3} we display  a representative case in which a deep
potential well, produced by choosing negative squared mass values for
$ m_{H_u}^2 $ and $ m_{\tilde L_i}^2 $,  gets removed  upon increasing
$\tilde \l _i /\l $.  We conclude that the lepton number violating
interactions can be effective for lifting the UFB-3 field direction.
\subsection{Electric Charge and Color   Breaking  Minima}

The lepton number violating interactions may generate a minimum of the
scalar potential along the field direction involving finite VEVs for
the charged Higgs-slepton fields, $ v_- = <H_d ^-> ,\ e_i = <\tilde e
_i > $.  This so-called type II charge breaking minimum has been
initially considered by Hirsch et al.~\cite{Hirsch:2004hr} for the
MSSM with bilinear R-parity violation.  We pursue a corresponding
analysis for the NMSSM with lepton number violation by solving the
equations of motion for the neutral fields in terms of the soft mass
parameters $m^2 _{H_u}, \ m^2 _{H_d}, \ m^2 _{S} $ and $m^2 _{\tilde
L_i} $ (corresponding to prescription II), and substituting these into
the two equations of motion for the charged fields, $ v_-$ and  $e_i $.
The equations are displayed in Eq.~(\ref{eomch}) of
Appendix~\ref{appexa}.  Note that we can ignore the VEV of the field $
\tilde e_i^c$ since $ e_i ^c =0$ appears to be the only solution for
$\l _i^e = 0$.  The equations for $ v_-$ and $e_i $ depend on the gauge
and Yukawa couplings, the soft mass parameters $ m ^2 _ {H_d\tilde
L_i}, $ and $v_i$.

To start with, we consider the limit of small lepton number violating
couplings obtained by expanding the equation of motion in powers of
the small parameters $\tilde \l _i $.  Substituting the expansion up
to leading order in $\tilde \l _i $ \bea && v_d = v_d^{(0)} + \tilde
\l _i v_d^{(1)} ,\ v_u = v_u^0 + \tilde \l _i v_u^{(1)} ,\ x = x^{(0)}
+ \tilde \l _i x ^{(1)} , \cr && v_i = \tilde \l _i v_i^{(1)} ,\ v_- =
\tilde \l _i v_- ^{(1)} ,\ e_i =\tilde \l _i e_i^{(1)} ,\eea in the
equations of motion for $ v_- $ and $e_i$, we find that the equations
in leading order of $\tilde \l _i $ only admit the trivial solution $
v_ - =0,\ e_i =0$.  Thus, we reach the conclusion that as long as the
couplings $\tilde \l _i $ are small compared to unity, the lepton
number violating interactions cannot cause the emergence of charge
breaking vacuum solution.  This conclusion for NMSSM corresponds to
that reached in Ref.~\cite{Hirsch:2004hr} for the MSSM.

To determine whether a non-trivial solution is favored at finite
$\tilde \l _i $, we attempt to solve numerically the equations of
motion for $ v_-$ and $ e_i, $ and compare the minimum value $
V_{min}$ with the regular minimum value $ V_{MIN}$ at a discrete set
of values of the parameters $ \tilde \l _i$ and $v_i$.  For fixed
$\tan \b$, we evaluate $V _{MIN} $ by using the formulas $ v_u = v'
\sin \b ,\ v_d = v' \cos \b , \ [ v' = (v^2 - v_i ^2) ^\ud] $.  Based
on the argument that the singlet VEV $x$ is not strongly affected by
the lepton number violating interactions, we identify the value of $
x$ in $V _{MIN} $ with its value along the charge breaking field
direction in $V _{min}$.  The values of $\tan \b $ near unity are
critical, as non-trivial solutions exist only around $ \tan\b =1$ and
disappear quickly at larger values.  Since the task of determining the
exclusion plot in the  $ \tilde \l _i , \ v_i$ plane is cumbersome, we
use typical values for the relevant parameters, so as to determine to
what extent a charge breaking solution at $ \l _i =0$ becomes
disallowed by increasing $ \l _i $. We restrict ourselves to solutions
with real values of $ v_-$ and $e_i$.  For the choice of input
parameters $\l =0.7,\ g_1^2=0.127,\ g_2^2= 0.425, \ \kappa =0.3, \ A
_\l = 500 \ \text{GeV} ,\ A _{\tilde \l _i } = 500 \ \text{GeV},\ A _
\kappa =250 \ \text{GeV},\ m^2 _{H_d \tilde L_i } = 100 \ \text{GeV}
^2$, with the fixed values of the VEVs $ x= 100 \ \text{GeV} $ and $
v_i = v/10 $, we find the real solution $(v _- , e_i) = (-60.9,
585.1)$ GeV having $ V_{min}- V_{MIN} \approx -8.26 \ \times  10 ^{9} \
\text{GeV}^4 $ for values of $ \tan \b =1 $ and $ \tilde \l_i /\l =0.1
$.  However, we find no non-trivial solutions as we increase the
coupling $\tilde \l _i /\l \in [ 0.5, 1] $.  We thus conclude that the
lepton number violating interactions have the ability to lift the
charge breaking minima.

Finally, we comment briefly on the issue of charge and color breaking
minima in the NMSSM by focusing on the field
directions~\cite{Frere:1983ag} described by $ v_u = \tilde u_i =
\tilde u^c_i =w$ and the singlet field VEV $x$, with all other fields
vanishing. Assuming $x$ to  be frozen, for simplicity, one finds that
the resulting potential, 
\bea &&  V _{CCB} = \kappa ^2 x ^4 + m_S ^2 x^2 -
{2\over 3} A_\kappa  \kappa x^3 + w ^2 [ 3 \l _i ^{u2} w^2 + 2 A ^u_i
\l ^u _i w + m^2 _{\tilde U_i} + m^2 _{\tilde U^c_i} + m^2 _{H_u}
+\sum _A (\tilde \l _A x )^2 ] , \eea does not develop a deep minimum
along $w$, for small Yukawa coupling constants $ \l ^u _i$, provided
one satisfies the conditions on the trilinear scalar matter couplings,
$ A _i^u \leq 3 ( m ^2 _{H_u} + m ^2 _{\tilde  U_i} + m ^2 _{\tilde U
^c_i} + (\tilde \l _A x ) ^2 ) ^ \ud $. 
This approximate  result suggests that the  bounds on $ A _i^u $
should  become  weaker upon increasing $ \tilde \l _i $.  
The  general field direction described by~\cite{Casas:1997ze} $
(\vert v_u \vert ,\ \vert u _i\vert ,\ \vert u^c_i\vert,\ \vert
v_d\vert ,\ \vert v_j\vert ) = (1, \a , \b , \g ,\ \g _L) w $ along
with the singlet field VEV $x$, involves a more elaborate discussion which we
shall not pursue here.

\section{\bf Conclusions} 
\label{sec5}

In the present work we have examined the effect of lepton number
conserving and violating Yukawa couplings of same size, $\tilde \l _i
\approx \l $, on NMSSM.  One expects significant modifications for the
scalar sector observables since the spontaneous electroweak gauge
symmetry breaking is now linked to both the down type Higgs boson and
sleptons.  An important challenge was raised by the need to define a
simple parameterization of the model consistent with the constraints
on the light neutrinos without specifying in detail the underlying
dynamics. Noting that the dominant contributions to the neutrino
Majorana mass matrix are controlled by the effective alignment
parameters $\tilde \l ' _i =\tilde \l _i - \l v_i /v_d $ and $ \eta _i
= { A _{\tilde \l _i} \tilde \l _i \over A_\l \l } - {v_i \over v_d}
$, we proceeded by treating $\tilde \l _i $ as free parameters while
determining the values of the sneutrino VEVs $ v_i$ and $ A_{\tilde \l
_i} $ through the restrictive conditions, $\tilde \l ' _i \simeq 0,\
\eta _i \simeq 0.$ There is no unique prescription, and the one used
may well single out a non-generic region of the NMSSM parameter space.
To answer this objection, one could attempt building a supergravity
unified model with a $U(1)_R$  symmetry spontaneously broken in a
hidden sector so as to check whether this respects an approximate
dynamical alignment robust under the renormalization group scale
evolution, following a similar analysis as that of Nilles and
Polonsky~\cite{nillespol97}.

In the first part of the present work we examined the impact of the
lepton number violating interactions on the mass spectra of scalars.
The vacuum stability constraints were found to restrict the VEV
parameters, $\tan \b = v_u /v_d $ and $ x$, to narrower intervals than
in the lepton number conserving case.  This property is reflected in
the fact that the bounds on $\tilde \l _i $ become stronger for larger
$\tan \b $ and $ x $.  There are certain analogies between our study
and that developed for the MSSM with bilinear R-parity violation by
Davidson et al.~\cite{davrius00} in terms of the basis invariant
parameter $\d _R$, corresponding to $ \sum _i \eta _i $ in our basis
choice, at fixed values of the CP-odd scalar masses.  The conclusions
in the latter work regarding the reduced range of variation of $\tan
\b $ and the size of the corrections to the scalar sector masses are
qualitatively similar to ours.  However, no meaningful comparison can
be made because of the different parameterizations.

The second part of the present work was devoted to a qualitative study
of the vacuum stability contraints from the UFB field directions and
the CCB minima.  The discussion for this case, unlike that in the
first part, does not bear directly on the implementation of the
parameter alignment conditions.  Based on illustrative examples
covering a small part  of the parameter space, we found that the
lepton number violating interactions may have a positive impact on the
regular vacuum stability provided the coupling constants $\tilde \l
_i$ and $ A _{\tilde \l _i} $ assume large enough values.  Since the
approach of selecting field directions based on the renormalization
group cannot be developed in the NMSSM by analytic means only, its
advantage over a systematic numerical exploration covering the full
field space is not clear.  Nevertheless, our discussion indicates that
pursuing the renormalization group approach on more quantitative
grounds is worthwhile.  The recent progress in developing efficient
numerical methods to search the global minimum of the scalar potential
in multidimensional field spaces~\cite{maniatis07} could  be useful
for further studies along these lines.

\appendix
\section{Useful Formulas and Conventions}\label{Notations}
\label{appexa} 

The $R$-parity conserving and $R$-parity violating parts of the NMSSM
superpotential can be written as \bea W _{RPC} & = & \l ^u _{jk} H_u
Q_j U_k^c + \l ^d _{jk} H_d Q_j D_k ^c + \l ^e _{jk} H_d L_j E_k^c +
\l H_d H_u S -{\kappa \over 3} S^3,
\label{nmssm1}\\
W _{RPV} & = & \ud \l _{ijk} L_iL_j E_k^c + \l '_{ijk} L_i Q_j D_k^c +
\ud \l ''_{ijk} U^c _i D^c_j D^c _k + \tilde \l _i L_i H_u S.
\label{nmssm2}
\eea The covariant four vector notation for the lepton and the
down-type Higgs superfields, and for the Yukawa couplings, employed in
this paper is \bea L_A & = & (H_d, L_i) = ({H_d ^0, \tilde H_d
^0\choose H_d^-, \tilde H_d^-} ,\ { \tilde\nu _i, \nu _i\choose \tilde
e_ i, e_ i } ),
\label{lepton1a}\\
\tilde\l _A & = & (\l , \tilde \l_i ),\ \l _{Ajk} = (\l ^e _{jk} , \l
_{ijk} ) ,\ \l '_{Ajk} = (\l ^d _{jk} , \l '_{ijk} ). \label{yukawa1a}
\eea We use the following convention in writing the multiplication of
chiral electroweak doublet superfields: \bea L_A H_u & \equiv & L_A
\cdot \e \cdot H_u = \nu _A H_u ^0 - e_A H_u^+ = (H_d ^0 H_u ^0 - H_d
^- H_u^+ ) + (\nu _i H_u ^0 - e_i H_u^+ ) ,\\ H_u Q & \equiv & H_u
\cdot \e \cdot Q = H_u ^+ D - H_u ^0 U,\ H_d Q \equiv H_d \cdot \e
\cdot Q = H_d ^0 D - H_d ^- U
, \eea where $\e $ is the $ 2\times 2 $ antisymmetric matrix with $ \e
_{12} =- \e _{21} =1$.  The $R$-parity conserving and $R$-parity
violating contributions to the trilinear part of the potential can be
written as \bea V _{RPC} ^ {soft} & = & - A_{jk} ^u \l_{jk} ^u H_u
\tilde Q_j \tilde U^c _k - A_{jk} ^d \l_{jk} ^d H_d \tilde Q \tilde
D^c _k - A_{jk} ^e \l_{jk} ^e H_d \tilde L_j \tilde E^c _k - A_{\l }
\l H_d H_u S - {A_{\kappa } \kappa \over 3} S^3, \nonumber \\ V _{RPV}
^ {soft} & = & - \ud A ^\l _{ijk} \l _{ijk}\tilde L_i\tilde L_j \tilde
E_k^c - A ^{\l '} _{ijk} \l '_{ijk} \tilde L_i\tilde Q_j \tilde D_k ^c
- \ud A _{ijk} ^ {\l ''} \l ''_{ijk}\tilde U^c _i \tilde D^c_j\tilde
D^c _k - A _{\tilde \l _i } \tilde \l _i \tilde L_i H_u S + H.\
c. . \eea The relationship that we adopted between the superpotential
and the effective Lagrangian uses the convention $ L_{EFF} = + [W]_F +
H.c. = - \vert W_i \vert ^2 +\ud W_{ij} \psi _i \psi _j + H.c. $ in
contrast to certain authors which use the opposite sign convention, $
L_{EFF} = - ([W]_F + H.c.)  = - \vert W_i \vert ^2 -\ud W_{ij} \psi _i
\psi _j + H.c. $.  The latter sign convention is that adopted, for
instance, in the studies of the NMSSM by Miller et
al.,~\cite{miller04} and by Barger et al.,~\cite{Barger:2006dh}.
Accounting for this fact, we obtain the following correspondence
between our notations and that of the latter authors: $\l \to h_s, \
A_ \l \to A_s,\ \kappa \to \kappa ,\ A_\kappa \to - A_\kappa ,\ \l ^u
\to h_t, \ A^u _t \to A_t ,\ x \to s/\sqrt 2 $.  We have compared our
formulas for the scalar sector potential and mass matrices and for the
neutralino sector and found complete agreement.

The scalar potential for the electrically neutral fields, with the
electrically charged and color non-singlet scalar fields set to zero,
is given by \bea V_F + V_D + V_{soft} & = & \vert \l v_d + \tilde \l
_i v_i \vert ^2 \vert x \vert ^2 + ( \vert \l v_u \vert ^2 + \vert
\tilde \l _i v_i \vert ^2 ) \vert x \vert ^2 \nonumber \\ & + & \vert
v_u (\l v_d + \tilde \l _i v_i )-\kappa x^2 \vert ^2 + G ^2_+ (\vert
v_u \vert ^2 -\vert v_d \vert ^2 -\vert v_i\vert ^2 ) ^2 \nonumber \\
& + &\bigg [ - A_\l \l v_d v_u x - A_{\tilde \l _i} \tilde \l _i v_i
v_u x - { A_\kappa \kappa \over 3} \vert x \vert ^3 + \mu ^2 _{Au} v_A
v_u + m ^2 _{H_d \tilde L_i} v_d v_i^\star + \ H.\ c. \bigg ]
\nonumber \\ & + & m^2 _{H_u} \vert v_u \vert ^2 + m^2 _{H_d} \vert
v_d \vert ^2 + m^2 _{\tilde L _i} \vert v_i\vert ^2 + m_S^2 \vert x
\vert ^2.
\label{potneut1} \eea

The relevant neutral and charged field VEVs are denoted as, $<H_u^0> =
v_u, \ <H_d^0> = v_d,\ <S>= x, \ <\tilde \nu _i > = v_i , $ and
$<H_d^-> = v_-, \ <\tilde e_i > = e_i ,\ <\tilde e ^c_i > = e_i ^c ,\
<\tilde q_i > = q_i ,\ <\tilde q ^c_i > = q_i ^c ,\ [\tilde q _i=
(\tilde u _i,\ \tilde d_i ) ,\ \tilde q ^c _i= (\tilde u ^c_i,\ \tilde
d ^c_i ) ].$ Our choice of field basis, $ v_+=0$, obviates the need to
consider the electrically charged direction $< H_u^+>= v^+ $.  The
formulas determining the soft masses of the Higgs bosons and sleptons
through the minimization equations of the neutral fields in
prescription I (using $ m^2 _{\tilde L_i }$ as free parameters) for
finite values of the neutral and charged fields are given by \bea &&
m^2 _{H_u} = {g_1^2\over 4} (e_i^2 +v_-^2 + v_d ^2 - v_u ^2 ) +
{g_2^2\over 4} (-e_i^2 -v_-^2 + v_d ^2 - v_u ^2 ) - \l ^2 v_d ^2 \cr
&& + \frac{ {A _\l \l }\, {v_d}\, {x}}{ {v_u}} - { {\l }}^2\,{ {x}}^2
- { {\tilde \l _i}}^2\,{ {x}}^2 + \frac{ {\kappa }\, {\l }\, {v_d}\, {
{x}}^2}{ {v_u}} \cr && + ( -2\, {\l }\, {\tilde \l _i}\, {v_d} +
\frac{ {A _{\tilde \l _i} \tilde \l _i }\, {x}}{ {v_u}} + \frac{
{\kappa }\, {\tilde \l _i}\,{ {x}}^2} { {v_u}} ) \, {v_i} + ( \frac{
{g_1 ^2}}{4} + \frac{ {g_2 ^2}}{4} - { {\tilde \l _i}}^2 ) \,{
{v_i}}^2 , \cr && \cr && m^2 _{H_d} = {g_1^2\over 4} (-e_i^2 -v_-^2 -
v_d ^2 +v_u ^2 ) + {g_2^2\over 4} (e_i^2 -v_-^2 - v_d ^2 + v_u ^2 )
\cr && - { {\l }}^2\,{ {v_u}}^2 + \frac{ {A _\l \l }\, {v_u}\, {x}}{
{v_d}} - { {\l }}^2\,{ {x}}^2 + \frac{ {\kappa }\, {\l }\, {v_u}\, {
{x}}^2}{ {v_d}} + {1\over v_d ^2 } \bigg [ ( - A _{\tilde \l _i}
\tilde \l _i v_u x - \kappa \tilde \l _i v_u x^2 ) \, {v_i} \cr && +
\bigg ( {g_1^2\over 4} (e_i^2 + v_-^2 -v_u^2 ) {g_2^2\over 4} (e_i^2 -
v_-^2 -v_u^2 ) + m^2 _{\tilde L _i } + \tilde \l _i ^2 (v_u^2
+x^2)\bigg ) v_i ^2 + {g_1^2 + g_2^2\over 4} v_i ^4 \bigg ] , \cr &&
\cr && m^2 _S = - { {v_-}}^2\,{ {\l }}^2 - 2\, {e_i}\, {v_-}\, {\l }\,
{\tilde \l _i} - { {e_i}}^2\,{ {\tilde \l _i}}^2 - { {\l }}^2\,{
{v_d}}^2 + 2\, {\kappa }\, {\l }\, {v_d}\, {v_u} - { {\l }}^2\,{
{v_u}}^2 \cr && - { {\tilde \l _i}}^2\,{ {v_u}}^2 + \frac{ {A _\l \l
}\, {v_d}\, {v_u}}{ {x}} + {A _\kappa \kappa }\, {x} - 2\,{ {\kappa
}}^2\,{ {x}}^2 + ( -2\, {\l }\, {\tilde \l _i}\, {v_d} + 2\, {\kappa
}\, {\tilde \l _i}\, {v_u} + \frac{ {A _{\tilde \l _i} \tilde \l _i
}\, {v_u}}{ {x}} ) \, {v_i} - { {\tilde \l _i}}^2\,{ {v_i}}^2 , \cr &&
\cr && m^2 _{H_d \tilde L_i } = - \ud {g_2 ^2}\, {e_i}\, {v_-} - {\l
}\, {\tilde \l _i}\, ( v_u^2 + x^2 ) + { 1\over v_d} \bigg [ \tilde \l
_i \, {v_u}\, x (A _{\tilde \l _i} + \kappa x ) \cr && + \bigg (
{g_1^2 \over 4} (-e_i^2 -v_-^2 -v_d^2 +v_u ^2 )+ {g_2^2 \over 4}
(-e_i^2 +v_-^2 -v_d^2 +v_u ^2 ) - m^2 _{\tilde L_i } \bigg ) v_i -
{g_1^2 + g_2^2 \over 4 } v_i ^3 \bigg ] . \label{eomneut} \eea

The equations of motion for the $L_A = (H_d , L_i) $ fields can be
expressed in the $SU(4)$ group covariant notation as \bea && \hat M^2
_{\tilde L, AB} v_B = \tilde \l _A x v_u (A _{\tilde \l _A } +\kappa x
) ,\cr && [\hat M^2 _{\tilde L, AB} = m^2 _{\tilde L, AB} + \tilde \l
_A\tilde \l _B (v_u ^2 + x^2) + 2 G_+ ^2 (\hat v_d ^2 - v_u ^2) \d
_{AB} ]. \eea

The above structure of the neutral scalars squared mass matrix
satisfies the important property that the alignment $\tilde \l
_A\propto v_A$ is satisfied if and only if $ \tilde \l _A$ is an
eigenvalue of the matrix $ \hat M^2 _{\tilde L, AB}.$ This result
generalizes that found in the MSSM with R-parity
violation~\cite{haberbasis,grossm04}.
 
The minimization equations for the charged slepton fields in
prescription II (using $ m^2 _{ H_d \tilde L_i }$ as free parameters)
are given by \bea && e^c _i: \ 0= 2\,{ {e^c_i}}^3\, {g_1^2} + { e ^c_i
}\, - [ { {e_i}}^2\, {g_1^2} + 2\, { m^2 _{\tilde E ^c} } - {g_1^2 }\,
( { {v_-}}^2 + { {v_i}}^2 + { {v_d}}^2 - { {v_u}}^2 ) ] - 2\, {A ^e
_\tau }\, {\l ^e _\tau } {e_i}\, {v_d} , \cr && \cr && v_-: \ 0=
{1\over {v_d}} \bigg [ { {e_i}}^2\, {g_2^2}\, {v_-}\, {v_d} + {v_-}\,
{v_u}\, ( {g_2^2}\, {v_d}\, {v_u} - 2\,{ {\l }}^2\, {v_d}\, {v_u} +
2\, {A_\l \l }\, {x} + 2\, {\kappa}\, {\l }\,{ {x}}^2 ) \cr && + 2\,
{e_i}\, ( {m^2 _{H_d \tilde L_i }}\, {v_d} + {\l }\, {\tilde \l _i }\,
{v_d}\,{ {x}}^2 )\bigg ] \cr && + {1\over v_d} \bigg [{ ( {e_i}\,
{g_2^2}\, ( -{ {v_-}}^2 + { {v_d}}^2 ) - 2\, {v_-}\, ( {m^2 _{H_d
\tilde L_i }} + {\l }\, {\tilde \l _i } \, ( { {v_u}}^2 + { {x}}^2 ) )
) \, {v_i}} \bigg ] - {g_2^2}\, {v_-}\,{ {v_i}}^2 , \cr && \cr && e_i:
\ 0= - {e_i \over {v_i}} ( {e_i}\, {g_2^2}\, {v_-}\, {v_d} + 2\, (
{m^2 _{H_d \tilde L_i }}\, {v_d} + {\l }\, {\tilde \l _i }\, {v_d}\, {
{v_u}}^2 - { A_{\tilde \l _i } \tilde \l _i }\, {v_u}\, {x} + {\l }\,
{\tilde \l _i }\, {v_d}\, { {x}}^2 - {\kappa }\, {\tilde \l _i }\,
{v_u}\, { {x}}^2 ) ) \cr && + {e_i}\, ( -2\,{ {\tilde \l _i }}^2\,{
{v_u}}^2 + {g_2^2}\, ( { {v_-}}^2 - { {v_d}}^2 + { {v_u }}^2 ) ) + 2\,
{v_-}\, ( {m^2 _{H_d \tilde L_i } } + {\l }\, {\tilde \l _i }\,{
{x}}^2 ) + {g_2^2}\, {v_- }\, {v_d}\, {v_i}
    \label{eomch}. \eea


\end{document}